\documentclass{aa} 

\usepackage{graphicx}
\usepackage{txfonts}
\usepackage{hyperref}
\usepackage{float}

\begin{document}

\title{Indications of an offset merger in Abell 3667}


   \author{Y. Omiya
          \inst{1}
          \and
          K. Nakazawa\inst{1,2}
          \and
          T. Tamura\inst{3}
          \and
          H. Akamatsu\inst{4,11}
          \and
          K. Matsushita\inst{5}
          \and
          N. Okabe\inst{6,7,8}
          \and\\
          K. Sato\inst{9}
          \and
          Y. Fujita\inst{10}
          \and
          L. Gu\inst{11,12,14}
          \and
          A. Simionescu\inst{11,12,13}
          \and
          Y. Ichinohe\inst{14}
          \and
          C. J. Riseley\inst{15,16}
          \and\\
          T. Akahori\inst{18}
          \and
          D. Ito\inst{1}
          \and
          K. Sakai\inst{1}
          \and
          K. Kurahara\inst{18}
          }

   \institute{Departure of Physics, Nagoya University, Furo-cho, Chikusa-ku, Nagoya, Aichi 464-8601, Japan\\
              \email{omiya\_y@u.phys.nagoya-u.ac.jp}
         \and
             Kobayashi-Maskawa Institute for the Origin of Particles and the Universe (KMI), Furo-cho, Chikusa-ku, Nagoya, Aichi 464-8601, Japan\\
             \email{nakazawa@u.phys.nagoya-u.ac.jp}
         \and
             Institute of Space and Astronautical Science, Japan Aerospace Exploration Agency, 3-1-1 Yoshinodai,Chuo-ku, Sagamihara, Kanagawa 229-8510, Japan
         \and
              International Center for Quantum-field Measurement Systems for Studies of the Universe and Particles (QUP), The High Energy Accelerator Research Organization (KEK), 1-1 Oho, Tsukuba, Ibaraki 305-0801, Japan
         \and
             Department of Physics, Tokyo University of Science, 1-3 Kagurazaka, Shinjuku-ku, Tokyo 162-8601, Japan
         \and
             Physics Program, Graduate School of Advanced Science and Engineering, Hiroshima University, 1-3-1 Kagamiyama, Higashi-Hiroshima, Hiroshima 739-8526, Japan
         \and
             Astrophysical Science Center, Hiroshima University, 1-3-1 Kagamiyama, Higashi-Hiroshima, Hiroshima 739-8526, Japan
         \and
            Core Research for Energetic Universe, Department of Physics, Hiroshima University, 1-3-1 Kagamiyama, Higashi-Hiroshima, Hiroshima 739-8526, Japan
         \and
             Department of Physics, Saitama University, 255 Shimo-Okubo, Sakura-ku, Saitama 338-8570, Japan
         \and
             Department of Physics, Graduate School of Science, Tokyo Metropolitan University, 1-1 Minami-Osawa, Hachioji-shi, Tokyo 192-0397, Japan
         \and
             SRON Netherlands Institute for Space Research, Utrecht, The Netherlands
         \and
             Leiden Observatory, Leiden University, PO Box 9513, 2300 RA Leiden, The Netherlands
         \and
         Kavli Institute for the Physics and Mathematics of the Universe (WPI), The University of Tokyo, Kashiwa, Chiba 277-8583, Japan
         \and
             RIKEN Nishina Center for Accelerator-Based Science, 2-1 Hirosawa, Wako, Saitama 351-0198, Japan
         \and
             Dipartimento di Fisica e Astronomia, Università degli Studi di Bologna, via P. Gobetti 93/2, 40129 Bologna, Italy
         \and
             INAF – Istituto di Radioastronomia, via P. Gobetti 101, 40129 Bologna, Italy
         \and
             Mizusawa VLBI Observatory, National Astronomical Observatory Japan, 2-21-1 Osawa, Mitaka, Tokyo 181-8588, Japan
             }

   \date{Received February XX, 2024; accepted March XX, 2024}

  \abstract
  {Cluster mergers are the most energetic events, releasing kinetic energies of up to 10$^{64}$~erg and involving megaparsec(Mpc)-scale shocks in their intra-cluster medium (ICM). In merging clusters, cold fronts are frequently observed, which are characterized by temperature and density jumps while maintaining constant pressure. They, together with the overall morphology of the ICM, provide important information for our understanding of the merging structure, such as velocity, impact parameter, and mass.}
  {Abell 3667 is a nearby ($z$=0.056) merging cluster with a prominent cold front and a pair of two bright radio relics. Assuming a head-on merger, the origin of the cold front is often considered to be a remnant of the cluster core stripped by its surrounding ICM. Some authors have proposed an offset merger scenario in which the subcluster core rotates after the first core crossing. This scenario can reproduce features such as the cold front and a pair of radio relics. To distinguish between these scenarios, we reanalyzed the ICM distribution and measured the line-of-sight bulk ICM velocity using the XMM-{\it Newton} PN data.}
  {We created an unsharp masked image to identify ICM features, and analyzed X-ray spectra to explore the ICM thermodynamical state. Applying the new XMM-{\it Newton} European Imaging Camera (EPIC)--PN calibration technique using background emission lines, the line-of-sight bulk ICM velocities were also measured.}
 {In the unsharp masked image, we identify several ICM features, some of which we detect for the first time. We confirm the cold front and note an enhanced region extending from the cold front to the west (named ``CF-W tail''). There is an enhancement of the X-ray surface brightness extending from the first brightest cluster galaxy (BCG) to the cold front, which is named the ``BCG-E tail''. The notable feature is a ``RG1 vortex'', which is a clockwise vortex-like enhancement with a radius of about 250~kpc connecting the first BCG to the radio galaxy (RG1). It is particularly enhanced near the north of the 1st BCG, which is named the ``BCG-N tail''. The thermodynamic maps show that the ICM of the RG1 vortex has a relatively high abundance of 0.5-0.6 solar compared to the surrounding regions. The ICM of the BCG-E tail also has a high abundance and low pseudo-entropy and can be interpreted as a remnant of the cluster core's ICM. Including its arc-like shape, the RG1 vortex supports the idea that the ICM around the cluster center is rotating, which is natural for an offset merger scenario. The results of the line-of-sight bulk ICM velocity measurements show that the ICM around the BCG-N tail is redshifted with a velocity difference of 940$\pm$440~km~s$^{-1}$ compared to the optical redshift of the first BCG. We obtain other indications of variations in the line-of-sight velocity of the ICM  and discuss these in the context of an offset merger.}
 {}


   \maketitle
%

\section{Introduction}





Cluster mergers are the most energetic events in the Universe since the Big Bang. During a merger, two (or more) galaxy clusters collide at a velocity of about 2000~km~s$^{-1}$, releasing energy amounting to as much as 10$^{64}$~erg \citep{2002ASSL..272....1S}. Through shock waves and turbulence in the intra-cluster medium (ICM), the gravitational potential energy is converted into ICM heating, accelerating the particles, and amplifying the magnetic field \citep[e.g.,][]{1970MNRAS.151....1W,1977ApJ...216..212J}. At the shock fronts, surface brightness edges and temperature jumps are observable via X-rays. Extended radio sources, called radio relics, are also observed
in the MHz-GHz range
(e.g., CIZA J2242.8+5301: \citet{2011MmSAI..82..569V,2013AN....334..342O}, Abell~2256: \citet{1996ApJ...465L...1M,2022ApJ...927...80R}, CIZA~J1358.9--4750: \citet{2023PASJ...75...37O,2023PASJ...75S.138K}, and Abell~3266: \citet{Riseley2022_A3266}). Several numerical simulations suggest that mergers generate shock waves with a velocity of 1000-2000 km~s$^{-1}$ at the shock front and 
accelerate particles with an efficiency of $\sim$10$^{-3}$ \citep[see][]{2018ApJ...857...26H}.
Observationally, application of the Rankine–Hugoniot equation to the X-ray temperature jump yields a shock velocity of this order of magnitude \citep[e.g.,][]{2004ApJ...606..819M,2013PASJ...65...16A,2023PASJ...75...37O}.

Cluster mergers also induce bulk motion and turbulence into the ICM. Simulations suggest that a ``major merger'', a nearly head-on merger of two large clusters, sometimes generates a large turbulent vortex spanning hundreds of kiloparsecs (kpc) and a velocity of $\sim$400~$\rm km~s^{-1}$ \citep[e.g.,][]{1999LNP...530..106N}. \citet{2013A&A...559A..78G} estimated a turbulent velocity of $\sim$500~$\rm km~s^{-1}$ in the Coma cluster using ICM density fluctuations based on {\it Chandra} and XMM-{\it Newton} observations.
The turbulence cascades down to smaller scales and eventually dissipates its kinetic energy,  thermalizing the ICM. Simultaneously, this turbulence can involve particle acceleration and magnetic field amplification \citep[e.g.,][]{2001MNRAS.320..365B,2001ApJ...557..560P,2003ApJ...584..190F,2005ApJ...622..205D,2007MNRAS.378..245B}, which are observed as radio halos in the MHz-GHz range \citep{2012A&A...543A..43V,2019SSRv..215...16V,2022A&A...660A..78B}.
These processes occur on long timescales of billions of years. Quantifying the motion  of the ICM is crucial for understanding the structure of cluster mergers and the amount of nonthermal energy of the ICM. In addition, the nonthermal pressure from the ICM motion and other nonthermal phases introduce a bias in the cluster mass estimate, which is a large systematic error in the cosmological gas and gravitational mass estimates \citep[e.g.,][]{2006MNRAS.369.2013R,2007ApJ...668....1N}.

In merging clusters, boundaries are frequently observed, characterized by temperature and density jumps while maintaining constant pressure. This is called a ``cold front'' and has been detected in some clusters by {\it Chandra} observations, such as Abell~2142, Abell~3667, and Abell~2146 \citep{2000ApJ...541..542M,2001ApJ...549L..47V,2010MNRAS.406.1721R}. The width of the cold front boundary in Abell~3667 is estimated in the {\it Chandra} image to be about 2-5~kpc, which is smaller than the typical mean free path of Coulomb scattering ($\sim$15~kpc). This indicates that particles crossing the boundary are suppressed by the magnetic field \citep[e.g.,][]{2001ApJ...551..160V}.
There are two possible origins of cold fronts. The first is the ``sloshing scenario'' \citep[e.g.,][]{2004ApJ...612L...9F,2006ApJ...650..102A,2007PhR...443....1M,2011MNRAS.413.2057R}, where the cold front is formed by the rotation of the low-entropy ICM in a perturbed gravitational potential of the cluster core due to the infall of the subcluster. In this case, the ICMs of the two regions on either side of the boundary originate from the same cluster.
The second is the ``stripping scenario'' \citep[e.g.,][]{2000ApJ...541..542M,2001ApJ...551..160V}, which forms the cold front when the ICM around the core of a subcluster is stripped by the tidal pressure as it falls into the main cluster. In other words, the boundary of the cold front is a ``contact discontinuity'' in the ICM across the two clusters. The shear flow of the ICM around the cold front stretches the initially tangled magnetic field lines to form a magnetic layer parallel to the front. Charged particles are trapped by the magnetic layer and orbited at a radius much smaller than the mean free path of Coulomb scattering \citep{2005AdSpR..36..636A}.

The Abell 3667 cluster is a prototypical X-ray-bright nearby merging cluster ($z$ = 0.056) in the southern sky. It has a pair of bright giant radio relics \citep[e.g.,][]{1997MNRAS.290..577R,2015MNRAS.447.1895R,2021PASA...38....5D} and a sharp cold front \citep{1999ApJ...521..526M,2001ApJ...551..160V}. {\it Suzaku} observations revealed the existence of a superhot (up to 20 keV) ICM \citep{2009PASJ...61..339N} and a shock associated with the northwest radio relic \citep{2012PASJ...64...67A}. The XMM--{\it Newton} observations confirmed this shock edge \citep{2010ApJ...715.1143F,2016arXiv160607433S} and the southeastern shock edge \citep{2018MNRAS.479..553S}. Recently, a giant halo extending from the northwest relic to the cold front was observed \citep{2022A&A...659A.146D}.
The {\it extended ROentgen Survey with an Imaging
Telescope Array (eROSITA)} showed the presence of a filament of 25–32~Mpc in length on the northwest side of the northwest relic \citep{2024arXiv240117281D}.

It has been proposed that the cold front of Abell~3667  originates from the stripping scenario \citep{2001ApJ...551..160V}. \citet{2017MNRAS.467.3662I} point out that the X-ray image near the cold front is similar to the image of the numerical simulation that modeled the ICM state of the inviscid stripping \citep{2015ApJ...806..104R}.
 \citet{2017MNRAS.467.3662I} concluded that the origin of the cold front is stripping from these features, and used the Kelvin-Helmholtz instability (KHI) at the boundary of the cold front to estimate the viscosity of the ICM. \citet{2006MNRAS.373..881P} showed that a 3:1 mass ratio merger reproduces the formation of two shocks moving in opposite directions, as well as a cold front \citep{2014ApJ...793...80D}. An extensive optical redshift survey by \citet[][]{2009ApJ...693..901O} suggested there are a few subgroups of galaxies based on Kaye's mixture modeling (KMM). These latter authors suggested that the southeastern group (KMM~5) is blueshifted with a velocity of $\sim $500~km~s$^{-1}$ compared to the cluster average value and that of the northwestern group (KMM~2). As the merging velocity is estimated to be 1500-2000~km~s$^{-1}$ based on the X-ray morphology and the temperature \citep[e.g.,][]{2002ASSL..272....1S}, 
\citet[][]{2009ApJ...693..901O} concluded that the merging axis is almost in the skyplane.

In contrast to the simple stripping scenario assuming a head-on merger, another possible formation mechanism of the cold front is line-of-sight flow.
Such motion can be generated by an ongoing merger assuming a particular geometry, such as an offset merger in a major merger,
or by sloshing caused by a minor merger in the past.
\citet{2016arXiv160607433S} first proposed an ``offset merger scenario'' to explain the location of the northwestern enhancement of the X-ray surface brightness, named the ``mushroom''. If there is a slight offset when two cluster cores pass each other, the subcluster core(s) begin to rotate orbitally and form a (pair of) cold front(s). These latter authors attributed the ``mushroom'' to being the counterpart of the cold front. In this scenario, the rotation-induced ICM flow and its velocity gradient shall exist. To distinguish between these scenarios, high-energy resolution microcalorimeter X-ray spectroscopy, such as those on board the X-Ray Imaging and Spectroscopy Mission ({\it XRISM}) \citep{2016SPIE.9905E..0VK,2020arXiv200304962X,2020SPIE11444E..22T,2023arXiv230301642S}, will be a powerful tool.

The average line-of-sight bulk velocity can also be measured using X-ray CCD data if sufficient Fe-K line statistics are available. Previously, bulk-motion measurements have been made using the CCD detector \citep[e.g.,][]{2014ApJ...782...38T,2016PASJ...68S..19O}. However, this method requires a precise calibration of the energy scale of the instrument. Recently, \citet{2020A&A...633A..42S} presented a revolutionary technique to measure the line-of-sight ICM velocities at the Fe--K complex lines with an accuracy of better than 150~km~s$^{-1}$ by recalibrating the energy scale in the European Imaging Camera (EPIC) PN detector on board the XMM--{\it Newton}. As the internal quiescent background emission lines (such as Cu-K and Ni-K$\alpha$) emitted from the base of the detector are used to recalibrate the energy scale, the central polygonal region covering 15\% of the field of view is not usable, however, by using a combination with multiple observations, it is possible to perform the bulk-velocity mapping in the bright and nearby clusters. Using this technique, \citet{2020A&A...633A..42S} measured the bulk velocities in the Coma cluster and the Perseus cluster. The ICM velocity in the latter cluster, estimated by applying their energy scale recalibration, is consistent with the results of the high-resolution X-ray spectroscopic observations made by the {\it Hitomi} satellite \citep{2016Natur.535..117H}, although not all of the fields of view are covered. For the Coma cluster, these latter authors found that the ICM velocity is consistent with the optical velocity in the central galaxies. In addition, the bulk-velocity measurements were performed for the Virgo cluster \citep{2022MNRAS.511.4511G,2023MNRAS.520.4793G}, the Centaurus cluster \citep{2022MNRAS.513.1932G}, and the Ophiuchus cluster \citep{2023MNRAS.522.2325G}.

In this paper, we first analyzed the ICM distribution of Abell~3667 using the long-exposure XMM--{\it Newton} PN data and then produced the bulk-velocity map by applying the gain recalibration method. The outline of this paper is as follows.
In Sec.~2, we describe two data-reduction methods, an image and spectroscopic analysis procedure, and a bulk-velocity measurement procedure.
In Sec.~3, we show, for the first time, the existence of a large vortex structure of about 250~kpc in radius, which we discovered using the standard XMM--{\it Newton} images, and discuss an offset merger scenario to explain it in Sec.~4. In Sec.~5, we show the newly obtained results of the bulk-velocity measurements applied to the central region using the gain recalibration method to place limits on or provide clues as to the offset merger scenario. 
In Sec.~6, we compared the radio image
of Abell 3667 at 943 MHz with our X-ray data. 
Throughout this paper, we assume $H_{0}~=~70~\rm{km~s^{-1}~Mpc^{-1}}$, $\Omega_{M}~=~0.3,$ and $\Omega_{\Lambda}~=~0.7$ (1~arcmin~=~65.64~kpc at $z$~=~0.056). Errors are given at the 68\% confidence ($1\sigma$) level unless stated otherwise.



\section{Data reduction}
\subsection{Observations}
\label{Appendix:recalibration)}

We used XMM--{\it Newton} observations pointing around the center of Abell~3667 as listed in Table~\ref{table2_Abell3667_obsid}. Standard processing of the observational data was performed using the Science Analysis Software (SAS) version 20.0.0 developed by the XMM--{\it Newton} Survey Science Center. The current calibration file (CCF) updated to May 2020 were used for comparison with the results of the gain calibration in \citet{2020A&A...633A..42S}. To mitigate the soft-proton effects, a time region was filtered using a good-time-interval (GTI) file with a threshold of 1~count~s$^{-1}$ in 100~s bins in the 10--15~keV energy band.

As mentioned above, two data-reduction processes are used in this paper: the ``image and spectroscopic analysis procedure" and the ``bulk-velocity measurement procedure". The former is a standard method commonly used for analyzing the ICM distribution, while the latter, proposed by \citet{2020A&A...633A..42S}, recalibrates the energy scale specifically for measuring the line-of-sight velocity of the ICM near the cluster core. This procedure improves the line energy calibration around the Fe-K lines but can only be applied to a subset of the total data.

\begin{figure}[htpb]
 \begin{center}
  \includegraphics[width=8cm]{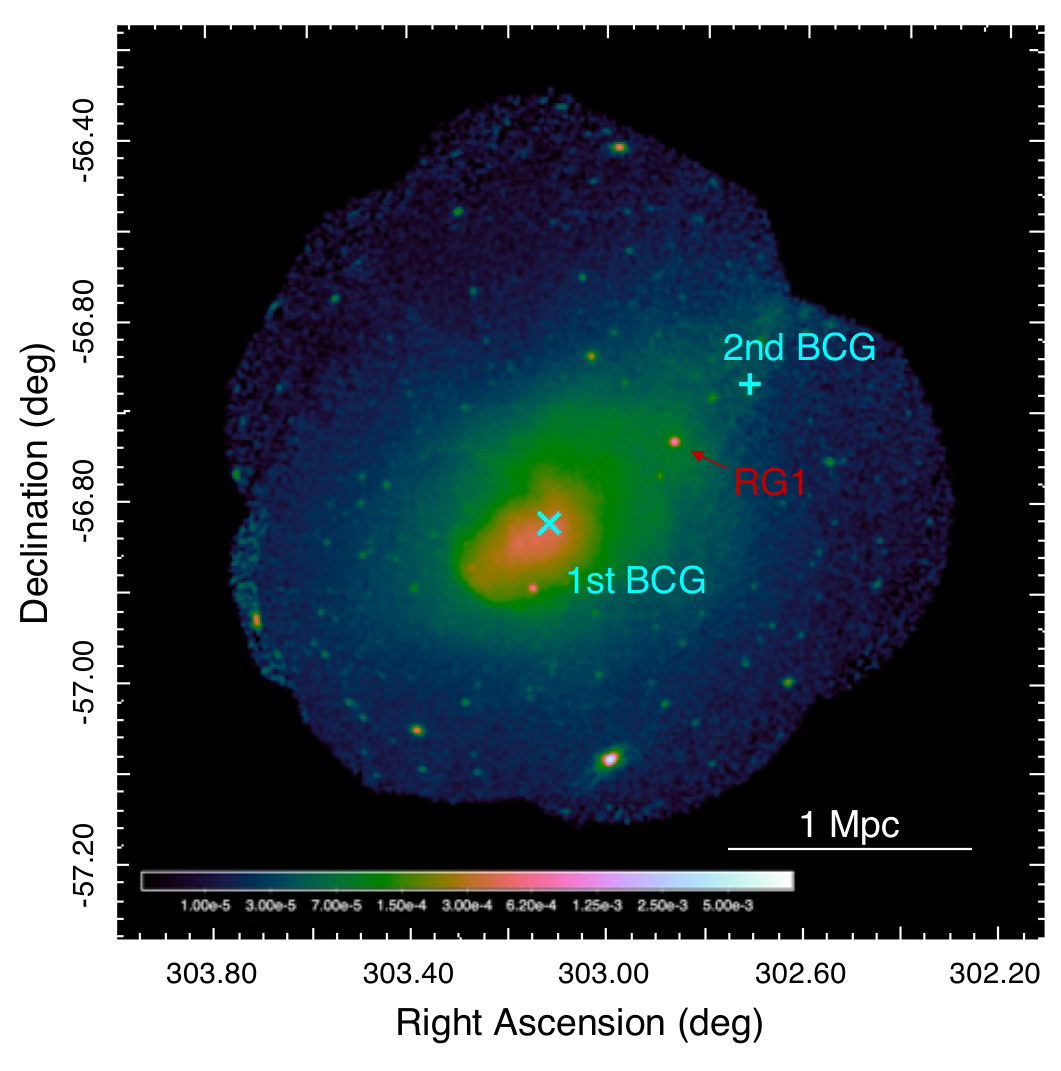}
 \end{center}
 \caption{Exposure-corrected and background-subtracted image in Abell~3667 in the 0.5--7.0~keV energy band. The unit is counts~s$^{-1}$~pixel$^{-1}$. The pixel size is 2$\arcsec\times$2$\arcsec$. The cyan ``x'' represents the position of the first BCG and the plus symbol marks the second.}
 \label{fig2.1_A3667_image}
\end{figure}

\subsection{Image and spectroscopic analysis procedure}

In the image and spectroscopic analysis procedure, the event data were calibrated following analysis procedures outlined in the XMM--{\it Newton} EPIC Cookbook\footnote{\url{http://heasarc.gsfc.nasa.gov/docs/xmm/esas/cookbook}}. A mosaic exposure-corrected and background-subtracted image in the 0.5--7.0~keV energy band is shown in Fig.~\ref{fig2.1_A3667_image}. Spectral and response files were generated using the {\it pn-spectra} task and the level of the quiescent particle background was estimated using the {\it pn-back} task. We utilized xspec v12.12.0 \citep{1996ASPC..101...17A} for spectral fitting. The thermal emission was modeled with the {\it apec} model with the abundance table in \cite{2009LanB...4B..712L}, and the redshift was fixed at 0.0556, except for the analysis of ICM velocity measurements. The hydrogen column density was fixed at $1.40\times10^{21}~\rm{cm^{-2}}$, determined from the HI map of the Leiden--Argentina--Bonn (LAB) survey \citep{2005A&A...440..775K}.
The sky background was decomposed into three components: the Local Hot Bubble (LHB), the Milky Way halo (MWH), and the cosmic X-ray background (CXB). These components were modeled following \cite{2023PASJ...75...37O}, with adjustments to the abundance table as per \cite{2009LanB...4B..712L}.

\begin{table}
\caption{Observation list in Abell~3667.}
\centering
\begin{tabular}{c|c|c|c}
    \hline \hline
    obsid & Duration & Revolution & mode \\
    \hline \hline
    0105260101 & 21~ks & 144 & EFF \\
    0105260201 & 20~ks & 155 & EFF \\
    0105260301 & 18~ks & 150 & EFF \\
    0105260401 & 17~ks & 150 & EFF \\
    0105260501 & 19~ks & 150 & EFF \\
    0105260601 & 26~ks & 149 & EFF \\
    0206850101 & 67~ks & 806 & EFF \\
    \hline
\end{tabular}
\label{table2_Abell3667_obsid}
\end{table}

\subsection{Bulk-velocity measurement procedure}
In the bulk-velocity measurement procedure, we followed the corrections in three steps  described by \citet{2020A&A...633A..42S}. The first-order correction is a time-dependent calibration of the average gain of the detector using the Cu-K$\alpha$ background line. The second-order correction is a detector-position-dependent calibration every 500 revolutions after the first-order correction has been applied. The third-order correction is an energy-scale calibration using several emission lines, including the Mn-K lines in the filter wheel closed (FWC) data illuminated by the $^{55}$Fe calibration source (hereafter CAL-FWC data). All of these data pointing to the center of Abell~3667 were obtained with the extended full frame (EFF) mode \citep{2001A&A...365L..18S} in the revolution of 0-1500 (2000$\sim$2008). Using observations matching these conditions, we estimated the functions of the three corrections needed to recalibrate the energy scale. The main difference from \citet{2020A&A...633A..42S} is our study uses CAL-FWC data with only the EFF mode, while \citet{2020A&A...633A..42S} used the full frame (FF) and EFF modes without distinction. As the gain recalibration focuses solely on correcting the energy scale around the Fe-K line, we employed this method exclusively for measuring bulk velocities. The details of the gain recalibration method are described in Appendix~\ref{section:SD20}.

\begin{figure}[htpb]
 \begin{center}
  \includegraphics[width=9cm]{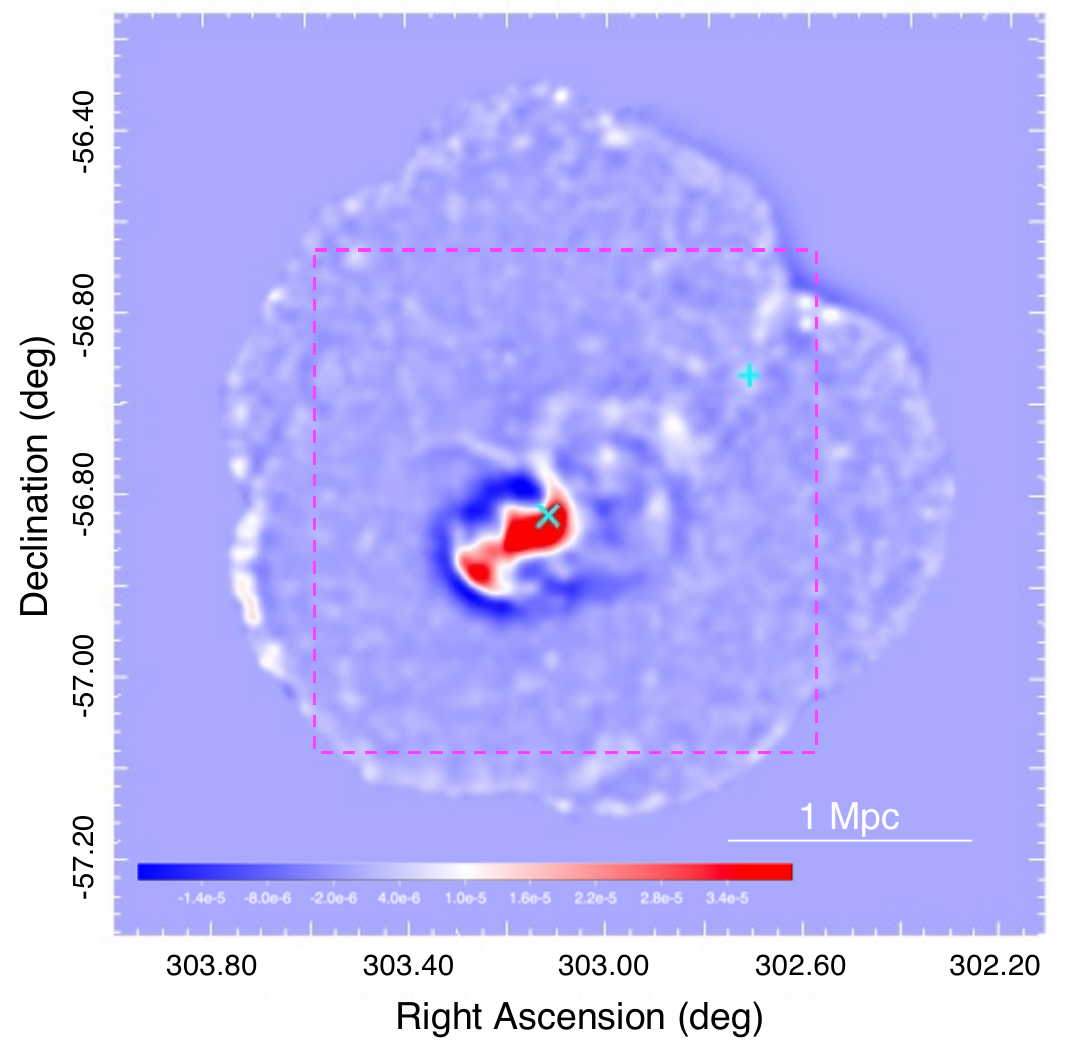}
 \end{center}
 \caption{Unsharp masked image created by subtracting images smoothed with $\sigma$=45$\arcsec$ and $\sigma$=180$\arcsec$. The outside of the field of view shows zero values. The pixel size is 2$\arcsec\times$2$\arcsec$. The magenta rectangle indicates a field of view in Fig.~\ref{fig3.1_Abell3667_image}.}
 \label{fig3.1_unsharp_image}
\end{figure}

\section{Features in ICM distribution in the central region}
\subsection{Tail and large vortex structures near the BCG}

\begin{figure*}[htp]
 \begin{center}
  \includegraphics[width=18cm]{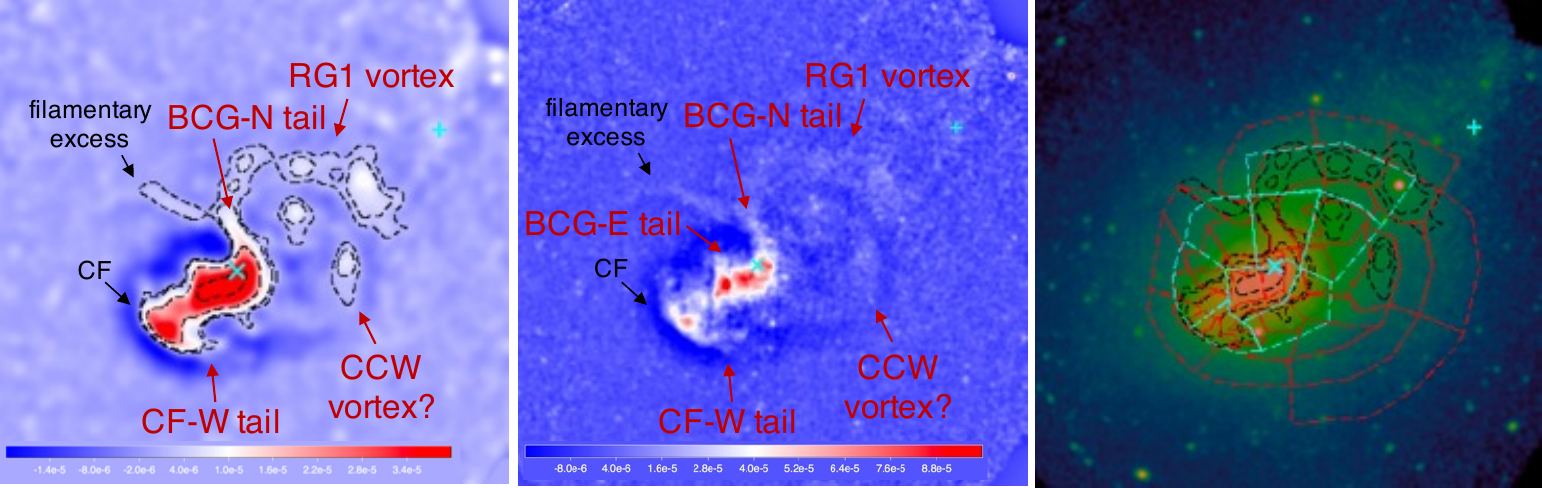}
 \end{center}
 \caption{
(Left) 
Close up of the cluster center. Same image as shown in Fig.~\ref{fig3.1_unsharp_image}, which is focused on the cluster center, with the field of view indicated by the magenta rectangle in Fig.~\ref{fig3.1_unsharp_image}.
It is overlaid with the black contours of 2, 10, 50, and 150$\sigma$ with respect to the rms noise.
(Middle) 
Unsharp masked image created by subtracting images smoothed with $\sigma$=3$\arcsec$ and $\sigma$=60$\arcsec$, presented to show fine structures. 
(Right)
Same image as shown in Fig.~\ref{fig2.1_A3667_image}.
Red polygons indicate the regions selected in the analysis of the thermodynamic and abundance maps discussed in Sec.~\ref{sec:Thermodynamic and abundance maps}.
Cyan polygons indicate the regions selected in the line-of-sight bulk-velocity measurements discussed in Sec.~\ref{sec_SD_result}.}
 \label{fig3.1_Abell3667_image}
\end{figure*}

In Fig.~\ref{fig2.1_A3667_image}, the locations of the first brightest cluster galaxy (BCG) and the second BCG are marked with a cyan ``x'' and a cyan plus symbol, respectively \citep{2004AJ....128.1558S,2011A&A...534A.109P}. The pixel size is 2$\arcsec\times$2$\arcsec$. To highlight the gradients, we created an unsharp masked image \citep[e.g.,][]{2006MNRAS.366..417F,2012A&A...539A..34D} by subtracting the X-ray surface brightness image smoothed with $\sigma$=180$\arcsec$ from that smoothed with $\sigma$=45$\arcsec$, as depicted in Fig.~\ref{fig3.1_unsharp_image}. Point sources with flux greater than 2.0$\times$10$^{-15}$~erg~cm$^{-2}$~s$^{-1}$ were identified and obscured by filling with random values from the surrounding region. The left panel of Fig.~\ref{fig3.1_Abell3667_image} presents the unsharp masked image focused on the cluster center overlaid with the contours at levels of 6.8$\times$10$^{-7}$, 3.4$\times$10$^{-6}$, 1.7$\times$10$^{-5}$, and 5.1$\times$10$^{-5}$~pixel$^{-1}$ or 2, 10, 50, and 150$\sigma$, with $\sigma$=3.4$\times$10$^{-7}$ pixel$^{-1}$ representing the image root mean square (rms) noise. The unsharp masked image indicates the existence of several special features in Abell~3667.

First, the location of the cold front is confirmed at the boundary between the enhanced and diminished regions to the southeast. Interestingly, the enhancement inside the cold front extends westward like a tail, spanning approximately 100~kpc in width. This ``CF-W tail'' edge also appears as a distinct step in a detailed analysis of the X-ray brightness distribution from {\it Chandra} \citep[see the 225$^{\circ}$–240$^{\circ}$ sector of Fig.~B1 in ][]{2017MNRAS.467.3662I}, marking the first clear identification of this feature to the best of our knowledge.

The filamentary excess extending from near the first BCG to the northeast has been noted in several papers, including \citet{2001ApJ...551..160V} and \citet{2017MNRAS.467.3662I}. 
In our unsharp masked image, the excess of this feature measures 2.1$\times$10$^{-6}$ pixel$^{-1}$ or 6.2$\sigma$. 
\citet{2002ApJ...569L..31M} first mentioned this feature and suggested that it provides evidence of colder, denser ICM in the cluster core being stripped out by hotter atmospheric ICM. Earlier studies  indicated that the filamentary excess has a relatively high Fe abundance, supporting the idea that it originates from the cluster core \citep{2009A&A...508..191L,2014ApJ...793...80D}.

In the unsharp masked image, we observed a previously unreported feature, a large clockwise vortex structure extending outward from the cluster center between the first and second BCGs.
The brightness is 7.6$\sigma$ or 2.6$\times$10$^{-6}$ pixel$^{-1}$. The vortex has a radius of about 250~kpc, and the structure seems to start at the first BCG. At the endpoint, there is a bright radio galaxy (B2007 -569) \citep[][]{1997MNRAS.290..577R,1999ApJ...518..603R,2015MNRAS.447.1895R}, which is named RG1 in \citet{2022A&A...659A.146D}.
The ``RG1 vortex'' is also visible in the XMM--{\it Newton} images in other papers \citep[e.g.,][]{2016MNRAS.460.1898S,2016arXiv160607433S,2018MNRAS.479..553S} and the {\it Chandra} image \citep{2017MNRAS.467.3662I}, although it has not been previously identified.
In particular, there is a clear enhancement near the first BCG of this feature, which we name the ``BCG-N tail''. We also identified a counterclockwise enhanced vortex (hereafter CCW vortex) structure with a radius of 250~kpc  extending from the cluster center outwards, with a significance of $\sigma$=3.3. However, due to the nature of unsharp masking and its relatively low significance, it is unclear whether this vortex represents the edge of the density jump and/or the inner layer of a ``dip'' inside the ``RG1 vortex''.

\subsection{Thermodynamic and abundance maps}
\label{sec:Thermodynamic and abundance maps}

\begin{figure*}[htp]
 \begin{center}
  \includegraphics[width=16cm]{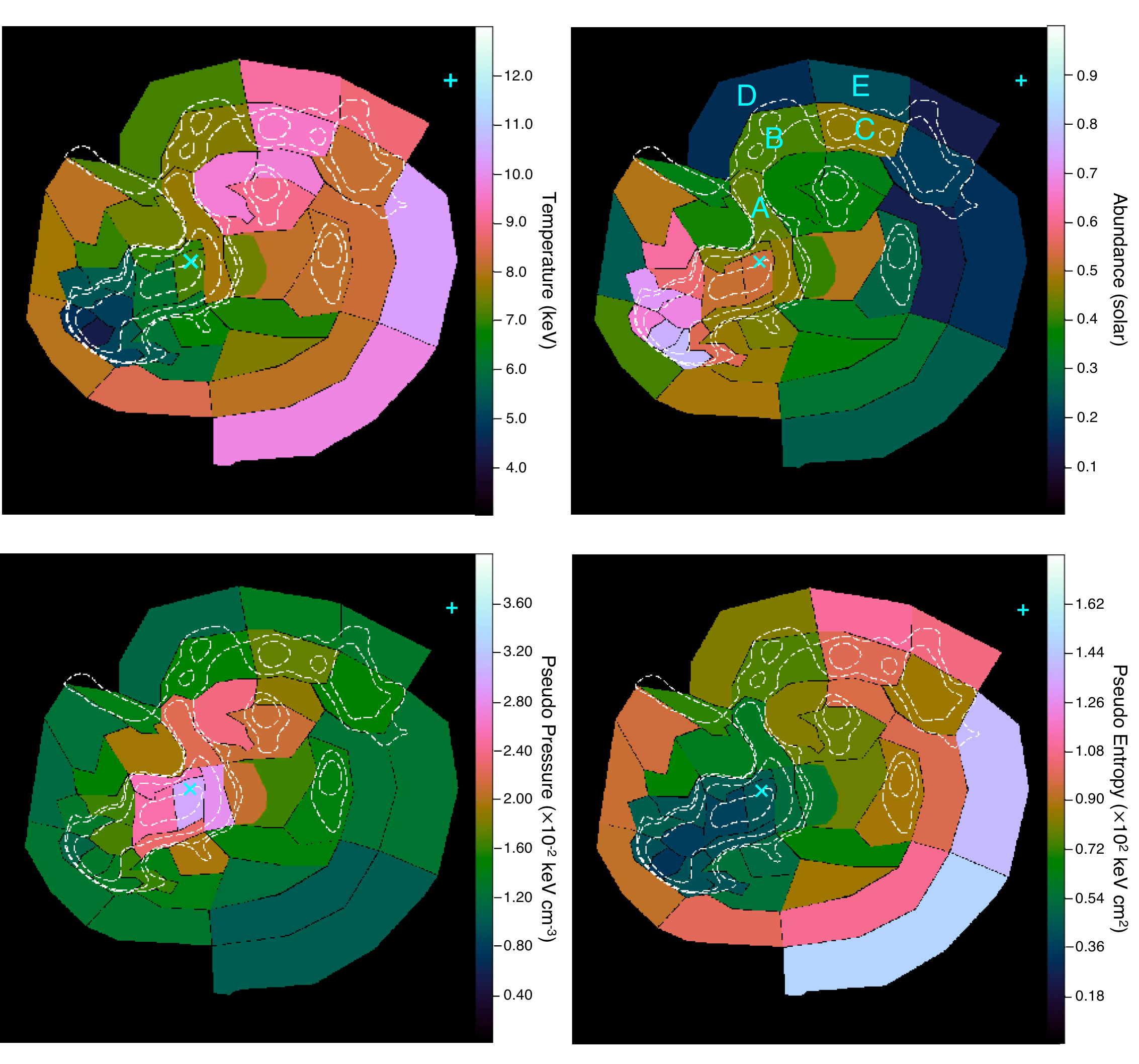}
 \end{center}
 \caption{Thermodynamic and abundance maps with regions as indicated by the red polygons in the right panel of Fig.~\ref{fig3.1_Abell3667_image}. The upper-left panel shows a temperature map overlaid with white contours that are the same as the black contours in the left panel of Fig.~\ref{fig3.1_Abell3667_image}.  The pseudo-density (upper right), pseudo-pressure (lower left), and pseudo-entropy (lower right) maps assume that the ICM is uniformly spread over the line-of-sight depth of 1~Mpc. The units of the 2D temperature, pseudo-density, pseudo-pressure, and pseudo-entropy maps are keV,  $\textrm{cm}^{-3}$ $\times$ $(l/1~\textrm{Mpc})^{-1/2}$,  $\textrm{keV~cm}^{-3}$ $\times$ $(l/1~\textrm{Mpc})^{-1/2}$ , and $\textrm{keV~cm}^{2}$ $\times$ $(l/1~\textrm{Mpc})^{1/3}$, respectively. The cyan symbols ("x" and "+") represent the BCGs, as in Fig.~\ref{fig2.1_A3667_image}.}
 \label{fig3.2_thermodynamic_map}
\end{figure*}

To investigate the large vortex structure, we generated two-dimensional thermodynamic and abundance maps. We delineated the regions along the enhancement or dip, as indicated by the red polygons in the right panel of Fig.~\ref{fig3.1_Abell3667_image}.
Figure~\ref{fig3.2_thermodynamic_map} shows 
the projected (or 2D)
temperature ($kT$), metal abundance ($\Lambda$), pseudo pressure ($\bar{P}$), and ``astrophysical entropy'' ($\bar{S}$) maps.
The $\bar{P}$ and $\bar{S}$ were calculated as $\bar{P}=\bar{n}_{\mathrm{e}}kT$ and $\bar{S}=\bar{n}_{\mathrm{e}}^{-2/3}kT$, respectively, using the pseudo density ($\bar{n}_{\mathrm{e}}$).
The $\bar{n}_{\mathrm{e}}$ was obtained from the normalization of the $\it{apec}$ model by assuming that the ICM is uniformly distributed over the line-of-sight depth of 1~Mpc, and that the electron density $\bar{n}_{\mathrm{e}}$ and the hydrogen density $\bar{n}_{\mathrm{H}}$ satisfy the relation $\bar{n}_{\mathrm{e}}=1. 2\bar{n}_{\mathrm{H}}$ considering the full ionization of helium.

\begin{figure}[htpb]
 \begin{center}
  \includegraphics[width=7cm]{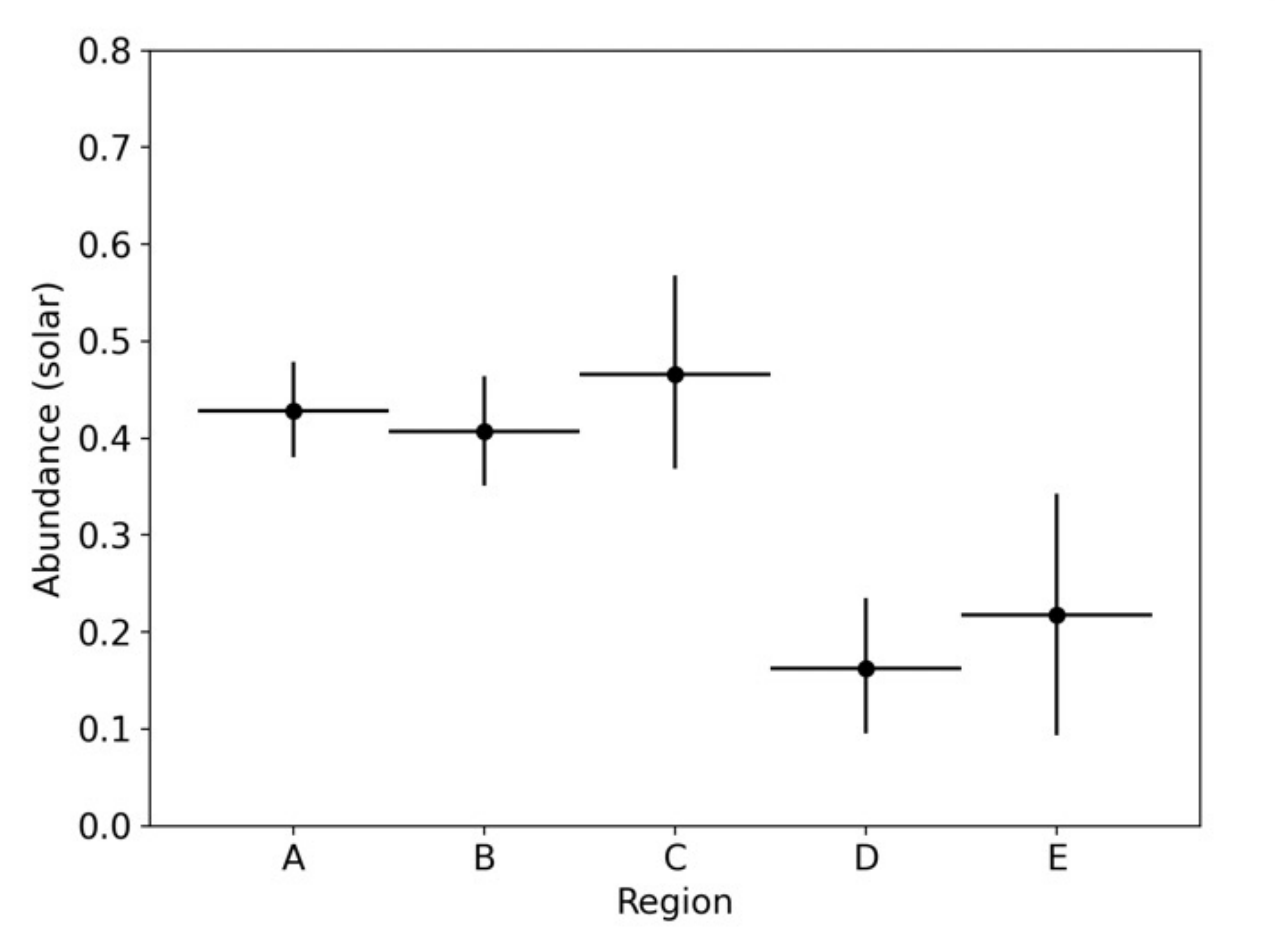}
 \end{center}
 \caption{Abundance profile in five~regions near the RG1 vortex with errors. The location of each region is indicated in the upper panel of Fig.~\ref{fig3.2_thermodynamic_map}.}
 \label{fig3.2_abund_profile}
\end{figure}

The RG1 vortex has high abundance comparable to the central region, with no significant pressure jumps at its boundaries. Figure~\ref{fig3.2_abund_profile} shows the results of the abundance in five regions near the RG1 vortex with error bars. Regions inside the vortex (regions A, B, and C; see also the upper right panel of Fig.~\ref{fig3.2_thermodynamic_map}) have significantly (a factor of about 2) higher  abundance than those outside (D and E).
These results suggest that the RG1 vortex structure is the remnant ICM of the cluster core.

\section{Possible scenarios for the ICM features}
\subsection{Karman Vortex scenario within a head-on merger}

Vortex structures have been identified in several clusters, but on a much smaller scale. The most prominent examples are KH rolls or vortices. Previous hydrodynamic simulations predicted that the KHI at the boundary of fluids with different densities and velocities would generate a 10-100~kpc-scale vortex in its shear layer \citep[e.g., ][]{2003MNRAS.346...13H,2010ApJ...717..908Z,2011MNRAS.413.2057R,2012A&A...544A.103V}. Indeed, the KH rolls have been observed in various clusters (e.g., Virgo: \citet{2016MNRAS.455..846W}, Abell~1775: \citet{2021ApJ...913....8H}, Abell~2142: \citet{2018ApJ...868...45W}, Abell~2319: \citet{2021MNRAS.504.2800I}). In Abell~3667, the KH rolls were also indicated at the boundary of the cold front using the deep {\it Chandra} observations. A plausible explanation for the presence of the large clockwise vortex is that these KH rolls developed within the inflowing stream where the subcluster core plunged.

Several fluid dynamics studies predict that a KH roll moves away from the shear layer, becomes a free vortex, and eventually forms a Karman vortex that aligns in a stable position \citep[e.g.,][]{2018JGRA..12310158C}. As the Reynolds number of the ICM is $\sim$~2100, the Karman vortex can be formed \citep{2017MNRAS.467.3662I}. In addition, because vortices of opposite sign are produced in each cycle, rows of alternating vortices are formed on the Karman vortex sheet. It is possible that the RG1 vortex and the CCW vortex flowing from near the BCG are located at the downstream end of the Karman vortex sheet.

The vortex shedding frequency is calculated to be 0.4~Gyr$^{-1}$, assuming that the Strouhal number is 0.2 \citep{2017MNRAS.467.3662I}. 
In Abell 3667, it is estimated that about 1~Gyr has passed after the collision \citep{2016arXiv160607433S}. In other words, theoretically, a few 250--500 kpc-sized vortices could potentially form. 
However, it remains unclear whether such large vortices can actually be formed within the timescale. An exploration of this scenario is deferred to a future study, as it necessitates a broad range of numerical fluid simulations, which is beyond the scope of this paper.

\subsection{Offset merger scenario}

\begin{figure*}[htpb]
 \begin{center}
  \includegraphics[width=16cm]{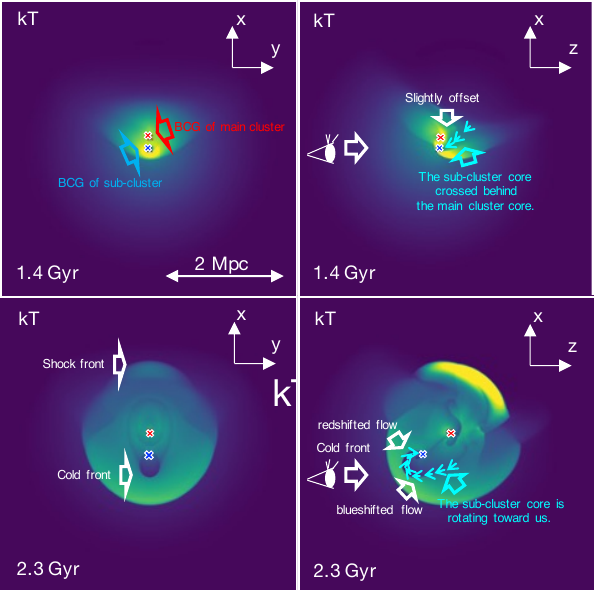}
 \end{center}
 \caption{Projected temperature distribution of an example case provided by the Galaxy Cluster Merger Catalog \citep{2016arXiv160904121Z,2018ApJS..234....4Z}. Time epochs after 1.4 Gyr and 2.3 Gy of collision initiation are shown. The initial parameter of the mass ratio is set to 3:1. The initial impact parameter is set at 1000~kpc in the z-axis direction, and the two clusters approach each other by moving along the x-axis. The locations of the BCGs in the subcluster and main cluster are indicated by blue and red crosses, respectively. Detailed results are shown in Fig.~\ref{fig3.3_zuhone_simulation} in Appendix~\ref{Detail of the offset merger simulation}.}
 \label{fig3.4_zuhone_simulation_pick}
\end{figure*}

There is little doubt that Abell~3667 is a nearly head-on major merger almost within the skyplane, as this scenario can explain the double radio relic and the relatively flat optical redshift distributions well. However, another possible scenario is that there was a small offset in the merger.
This scenario is also addressed in \citet{1999ApJ...518..603R} and \citet{2016arXiv160607433S}.
In such an offset merger, it is natural for the ICM to acquire angular momentum, leading to the production of vortex structures as the clusters interact and strip each other.

\subsubsection{Comparison with simulations}
\label{Comparison with simulation}

We compared our results with merger simulations provided by the Galaxy Cluster Merger Catalog\footnote{\url{http://gcmc.hub.yt}} \citep{2016arXiv160904121Z,2018ApJS..234....4Z}.
Figure~\ref{fig3.4_zuhone_simulation_pick} shows simulated physical parameter distributions of the ICM of a 3:1 mass cluster merger,
which is the same mass ratio condition for Abell~3667 as that presented in \citet{2006MNRAS.373..881P} and \citet{2014ApJ...793...80D}.
The initial impact parameter is set to 1000~kpc in the z-axis direction (see the panels at 0~Gyr in Fig.~\ref{fig3.3_zuhone_simulation}). The two clusters approach each other, moving along the x-axis as time passes.
The simulation results show that at 1.4~Gyr there is a slight offset in the center of the cluster cores as they cross each other. The ICMs of the two clusters are stripped from the surrounding high-temperature ICMs. The remnants of the cluster cores then rotate due to angular momentum. As the two BCGs (or the central cores of the cluster's dark matter halo) begin to move back towards each other for the second encounter, the ICMs will also be pulled back following the BCGs (see Fig.~\ref{fig3.3_zuhone_simulation}).

As shown in Fig.~\ref{fig3.3_zuhone_simulation}, the simulation results  show that an offset merger scenario can easily produce large ICM vortex structures. Their locations do not coincide with the location of the RG1 vortex, but this is reasonable because the simulation is not specifically performed to reproduce that of Abell 3667.
At 1.9--2.3~Gyr, when the subcluster core is rotating, the results in the y-z plane (and the x-y plane) show that the ICM of the subcluster bulges out during rotation and its front is observed as the sharp X-ray brightness and temperature jumps at the cold front position on the skyplane (see Fig.~\ref{fig3.4_zuhone_simulation_pick}). Therefore, the existence of the sharp cold front and the relics in Abell~3667 can also be explained by an offset merger scenario.
The rotation of the high-density ICM on the sky plane appears as a ``slingshot'' structure, which is a phenomenon that has been observed in several merging clusters (e.g., the Coma cluster: \citet{2022ApJ...934..170L}; Abell~2061:~\citet{2014xru..confE.181S}, ZwCl 2341+0000;~\citet{2021A&A...656A..59Z}).

This scenario is also supported by the location of the first BCG.
The first BCG is about 500~kpc behind the cold front in the sky plane, as shown in Fig.~\ref{fig2.1_A3667_image}.
In Fig.~\ref{fig3.4_zuhone_simulation_pick} and Fig.~\ref{fig3.3_zuhone_simulation}, the locations of the BCGs in each cluster are marked with blue (first BCG) and red (second BCG) crosses. As shown in Fig.~\ref{fig3.3_zuhone_simulation}, up to 1.9~Gyr, the first BCG is associated with the cold front, but then the BCG is pulled back by the gravitational potential of the other cluster center.
Meanwhile, the cold front is still pushed forward by the bulging ICM of the subcluster. Therefore, an offset merger scenario can explain that the BCG was about 500~kpc behind the cold front in the sky plane for Abell~3667.
Such a feature is also seen in another double-radio-relic merger, Abell~3376, where the BCG was about 170 kpc behind the cold front \citep{2013A&A...560A..78D,2018A&A...618A..74U}.

\subsubsection{X-ray tail structures observed around the BCG}
In Sec.~3.1, we identified some tail structures around the first BCG, including the BCG-N tail.
The unsharp masked image focused near the cold front (middle panel of Fig.~\ref{fig3.1_Abell3667_image}) shows
that there are two tail structures associated with the first BCG. The BCG-E tail to the southeast, and the BCG-N tail to the north. The enhancement of the BCG-E tail corresponds to $\sim$~10\% of the original surface brightness.
The Gaussian gradient magnitude (GGM)-filtered image also shows the existence of a similar structure \citep[see Fig.~4 in][]{2016MNRAS.460.1898S}. Assuming that the first BCG is moving from the cold front to the center, the enhancement appears to trace the movement of the first BCG. One possibility is that this is the interstellar medium (ISM) (or the remnant of the ICM in the subcluster core) being stripped by the surrounding ICM.

The BCG-E tail can be explained by the remnant ICM of the subcluster core as well as the BCG-N tail (part of the RG1 vortex).
The abundance map in Fig.~\ref{fig3.2_thermodynamic_map} shows that the ICMs in these regions are 0.4--0.6 solar, which is about 1.5 times higher than those of the surrounding regions. The pseudo-entropy values are two to three times lower than the surrounding ICM, and comparable to those near the cold front. These are the features seen in the cluster core.

\section{Measuring the line-of-sight bulk velocity with the gain recalibration method}
\label{sec_SD_result}

The redshift of the first BCG is 0.0556, while the redshift of the second BCG is 0.0560 \citep{2004AJ....128.1558S},  almost the same value, with a velocity difference of 120~$\pm$~86~km~s$^{-1}$. This leads us to ask the question of how an offset merger scenario can explain the coincidence of the two BCG line-of-sight velocities and their apparent alignment to the merger axis. If the cores of the two clusters crossed and rotated almost on the sky plane, such as in the case of merging clusters with the slingshot structure, the redshift coincidence can be easily explained, but not the alignment. Another possible scenario to explain the coincident redshift of the BCGs may be formed when the individual BCGs are near the apogee of their orbital rotation after the first crossing. As the orbital plane will be nearly on the line of sight, the apparent alignment of the orientation on the sky plane with the merger axis can also be explained. In that case, the ICM velocity in the line of sight would be dispersed because the ICM flow would remain. Therefore, we measured the bulk-velocity distribution in the line of sight using the gain recalibration method.

\subsection{Overall ICM redshift distribution of Abell 3667}

\begin{figure*}[htpb]
 \begin{center}
  \includegraphics[width=16cm]{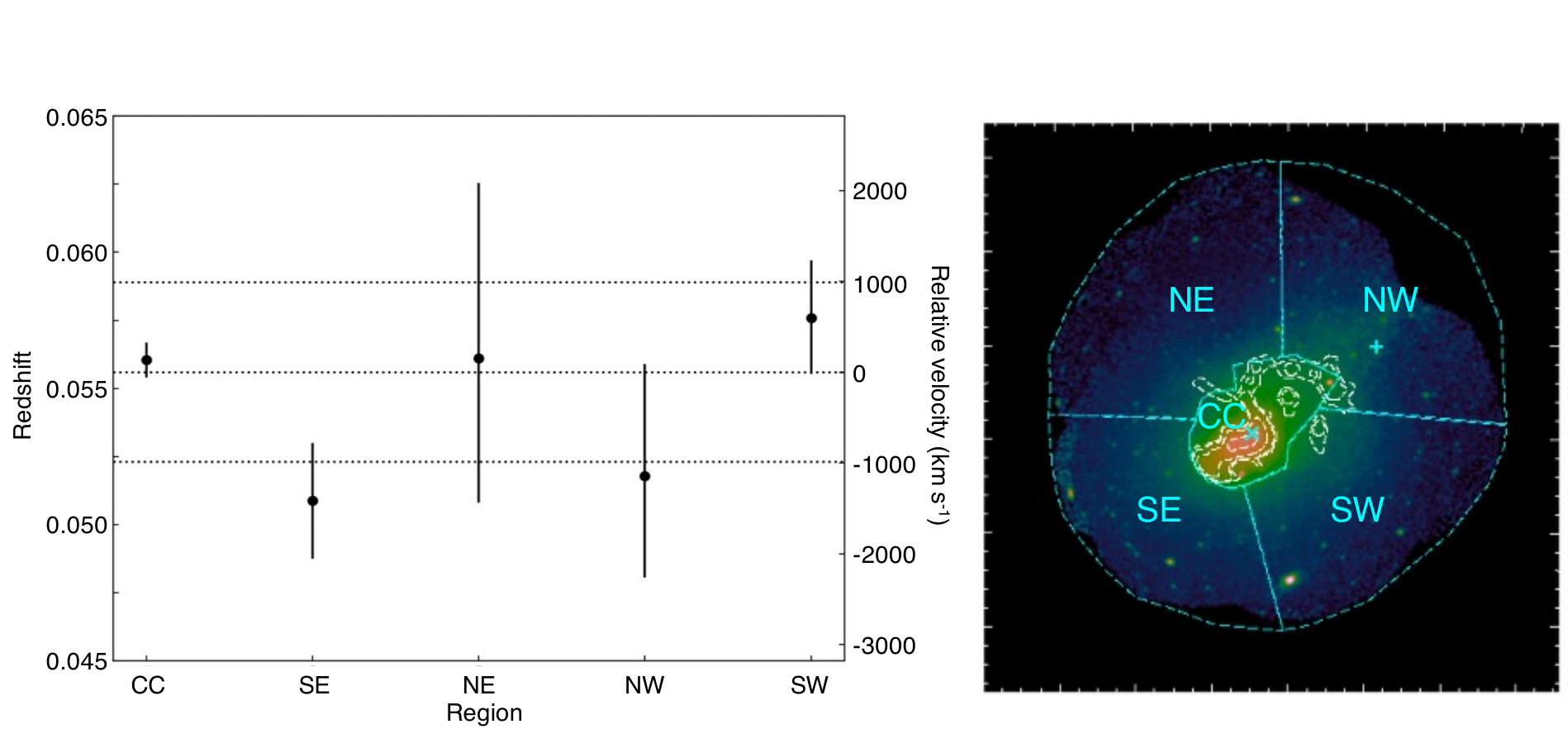}
 \end{center}
 \caption{Redshift measurements in the large regions. The region we define is overlaid on the right panel, which is the same image as in Fig.~\ref{fig2.1_A3667_image}. The horizontal axis is the region number given in the right panel of Fig.~\ref{fig5.2_velocity_results}. The right-hand scale in the redshift distribution is the relative velocity based on the optical redshift of the first BCG ($z$=0.0556). As a guide, the velocity differences of $\pm$1000~km~s$^{-1}$ are shown as the two dotted lines.}
 \label{fig5.3_large_velocity_results}
\end{figure*}

Before discussing the details of the central portion, we present the large-scale velocity structure obtained using the gain recalibration method. As shown in the right panel of Fig. \ref{fig5.3_large_velocity_results}, the entire field of view of the data we used was divided into five large regions: one at the cluster center and the others surrounding it. The results are shown in Fig.~\ref{fig5.3_large_velocity_results}.

The redshift of region~CC, located at the center of the cluster, is consistent with the optical redshift of the first BCG with considerable accuracy. Furthermore, the redshifts of regions~NE, NW, and SW are also consistent with the redshift of the first BCG within a 1$\sigma$ confidence interval. The consistency of these velocities suggests that the overall merger is taking place close to the skyplane. It should be noted, however, that the measured redshifts are not well constrained due to the low surface brightness.

On the other hand, region~SE tends to be blueshifted compared to region~CC. The relative velocity difference with respect to the optical redshift of the first BCG is $-$1400~$\pm$~630~km~s$^{-1}$, indicating a deviation of 2.3$\sigma$. This region is located outside the cold front and along the merger axis. Such blueshifts can be explained by a marked asymmetric flow in the case of an offset merger on the line of sight, or alternatively, by the inclination of the merger axis in a head-on merger scenario. It is therefore difficult to distinguish between these scenarios in the overall redshift distributions of the ICM. We note that region~SE coincides with the subgroup of galaxies named KMM~5 in \cite{2009ApJ...693..901O}, which has a redshift of about $-500$~km s$^{-1}$ (i.e., blueshifted).

\subsection{Central ICM redshift structures in the cluster center}

\begin{figure*}[htpb]
 \begin{center}
  \includegraphics[width=18cm]{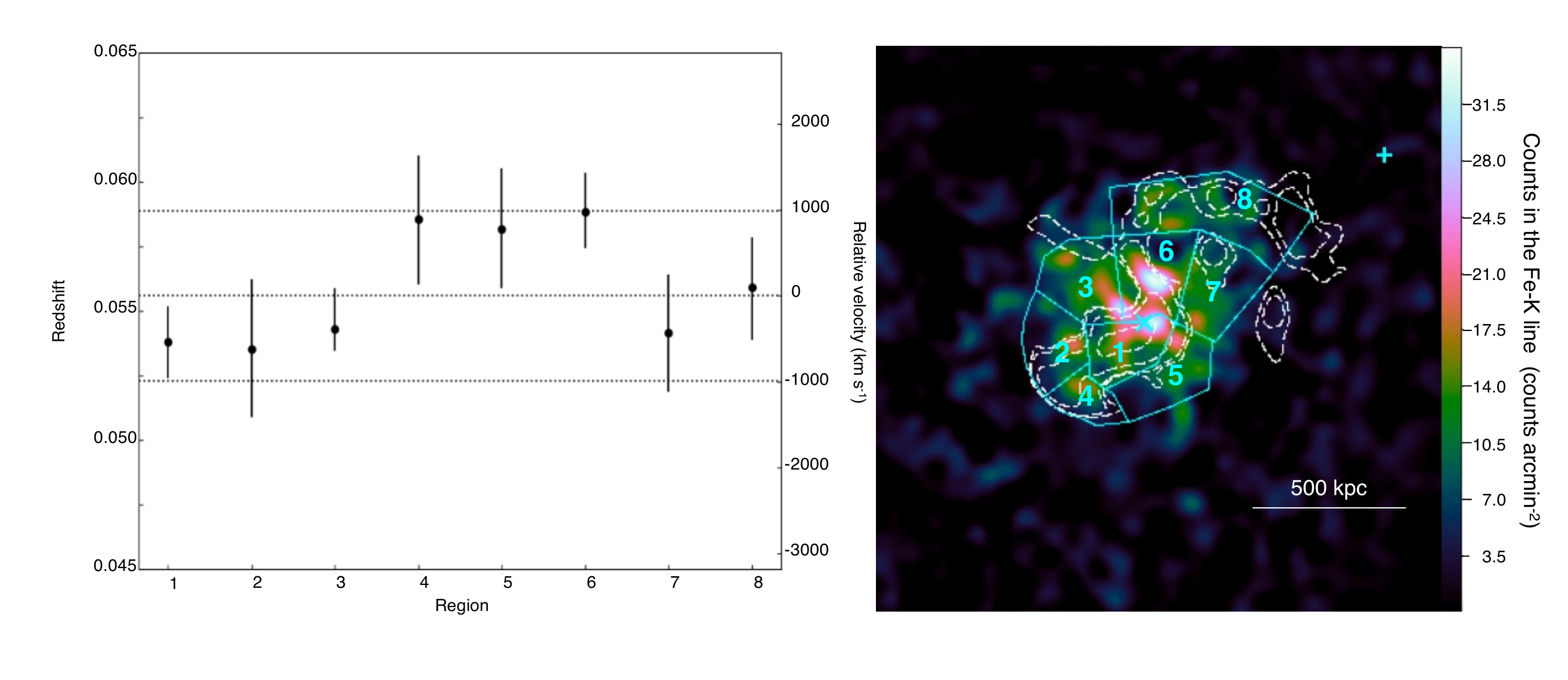}
 \end{center}
 \caption{Redshift measurements in the small regions. (Left) Redshift distribution. The horizontal axis is the region number indicated in the right panel of Fig.~\ref{fig5.2_velocity_results}. The right-hand scale in the redshift distribution is relative velocity based on the optical redshift of the first BCG ($z$=0.0556). As a guide, the velocity differences of $\pm$1000~km~s$^{-1}$ are indicated as the two dotted lines. (Right) Count map in the FeK line. Cyan polygons indicate the regions selected for the line-of-sight bulk-velocity measurements. The white contours are the same as the black contours in Fig.~\ref{fig3.1_Abell3667_image}.}
 \label{fig5.2_velocity_results}
\end{figure*}

As the next step in our approach to measuring the line-of-sight bulk velocity of Abell 3667, we examined region~CC in detail. Because its statistics are high, it can be divided into several subregions.
To properly select the regions, we created the Fe-K count map. The energy range of the emission line is defined as 6.4--7.1~keV. We also define band~2 (5.7--6.4~keV) and band~3 (7.1--7.8~keV) in the same energy interval on the low- and high-energy sides individually. The continuum component in band~1 was reproduced by the average counts of band~2 and band~3, and the counts of the Fe-K emission line were calculated by subtracting the average counts from the counts of band~1.
The ranges in bands~1-3 were shifted by multiplying by 1+$z$. 
The results are shown in the right panel of Fig.~\ref{fig5.2_velocity_results}.
As indicated by the cyan polygons, we subdivided eight regions so that the Fe-K emission line counts amount to 50--200 counts in each region.

The results of the line-of-sight bulk ICM velocity measurements in individual regions are shown in the left panel of Fig.~\ref{fig5.2_velocity_results}.
The redshift in region~1, which covers the southeastern half of the first BCG, is consistent with that of the first BCG within a 1.3$\sigma$ confidence interval. The redshifts in the regions between the cold front and the first BCG (reg~1-5) are also individually consistent within the 1.2$\sigma$ confidence interval. Here we performed the chi-squared test assuming that the redshifts of the ICM in all regions coincide with the redshift of the first BCG. The p-value was estimated to be 0.15, which is consistent with this scenario.

However, if we look at the individual values, 
it is noticeable that the line-of-sight velocity in region~6 is redshifted by 980~$\pm$~440~km~s$^{-1}$ and differs from that of the first BCG by 2.3$\sigma$ (or by $1.3$\% upper-tail probability). The region is located north of the BCG and is connected to the RG1 vortex. In other words, the BCG-N tail shows an indication of redshift.

From the left panel of Fig.~\ref{fig5.2_velocity_results}, we can see that regions~2 and 3, which are located to the east of region~1 (the head of the cold front), show a tendency to be blueshifted, while regions~4 and 5, which are located to the southwest, tend to be redshifted. To improve the statistics, we performed a simultaneous fitting by merging the spectra of regions 2 and 3  and those of regions 4 and 5. In this case, the parameters of temperature and abundance were free in individual regions, and the parameter of redshift was constrained to have the same value. The redshift was estimated to be 5.40$_{-0.14}^{+0.19}\times$10$^{-2}$ in regions~2 and 3 and 5.84$_{-0.17}^{+0.17}\times$10$^{-2}$ in regions~4 and 5. The velocity difference between the two group of regions is 1300~$\pm$~860~km~s$^{-1}$. The p-value, assuming that their redshifts are the same as the redshift of the BCG, is estimated to be 0.03.
In summary, from the perspective of the cold front (region 1) looking southeast, the ``right side'' is blueshifted and the ``left side'' is redshifted.

\subsection{Comparison with simulations for discussion of the merging picture}

Simulation results at 2.3~Gyr in Fig.~\ref{fig3.4_zuhone_simulation_pick}, when the BCGs are close to apogee, show that the ICM of the central cluster rotates in an arc, bulging outward. The predicted motion of the ICM of the  subcluster
core is indicated by the cyan arrows. If we were looking from the ``$-$z'' direction, the ICM would be blueshifted in the region near the cold front (region~1). Also, the arc-like motion of the ICM in the cluster core falling back toward the BCG (region~6) would appear redshifted. If this interpretation is correct, then the subcluster core crossed behind the main cluster core and moved toward us, as shown in the upper right panel of Fig.~\ref{fig3.4_zuhone_simulation_pick}.

In this case, the BCG-N tail would appear as an enhancement of the backward motion of the ICM due to the rotation of the subcluster core. On the other hand, the BCG-E tail is likely the stripped-off ISM of the first BCG that has become detached during the rotation.
The redshifts of regions~2 and 3 and 4 and 5, symmetrically located on the collision axis, could be attributed to the rotation of the subcluster, whose axis of rotation is slightly tilted with respect to the line of sight.

Although several regions with significant line-of-sight velocity shifts were identified, our ICM velocity analysis based on XMM--{\it Newton} data does not provide clear results enabling us to unambiguously identify the merger scenario.
Therefore, {\it XRISM} observations with accurate bulk and turbulence velocity measurements are crucial. Velocity mapping with multiple observations around the cold front and BCG would not only help distinguish the merger scenario in Abell~3667, but would also advance our understanding of ICM physics, such as viscosity and turbulence generation.

\section{Comparison with ASKAP radio observation}
In recent years, many extended radio sources have been newly detected using radio telescopes like MeerKAT \citep{2016mks..confE...1J} and the Australian Square Kilometre Array Pathfinder (ASKAP) \citep{2011PASA...28..215N}. 
Morphology of the cluster radio halo reflects the merger geometry.
Here, high sensitivity ---in order to detect largely diffuse continuum radio sources--- and higher angular resolution ---in order to identify contributions from point sources--- are both essential.

The project of Evolutionary Maps of the Universe (EMU) with ASKAP is a pilot radio continuum survey of the southern sky. It provides images up to a sensitivity of $\sim$20 $\mu$Jy beam$^{-1}$ rms with a beam size of $15''\times 15''$ 
at 943~MHz with an instantaneous 288~MHz bandwidth. These characteristics are ideal for cluster radio halo and relic morphology surveys.

\begin{figure}[htpb]
 \begin{center}
  \includegraphics[width=9cm]{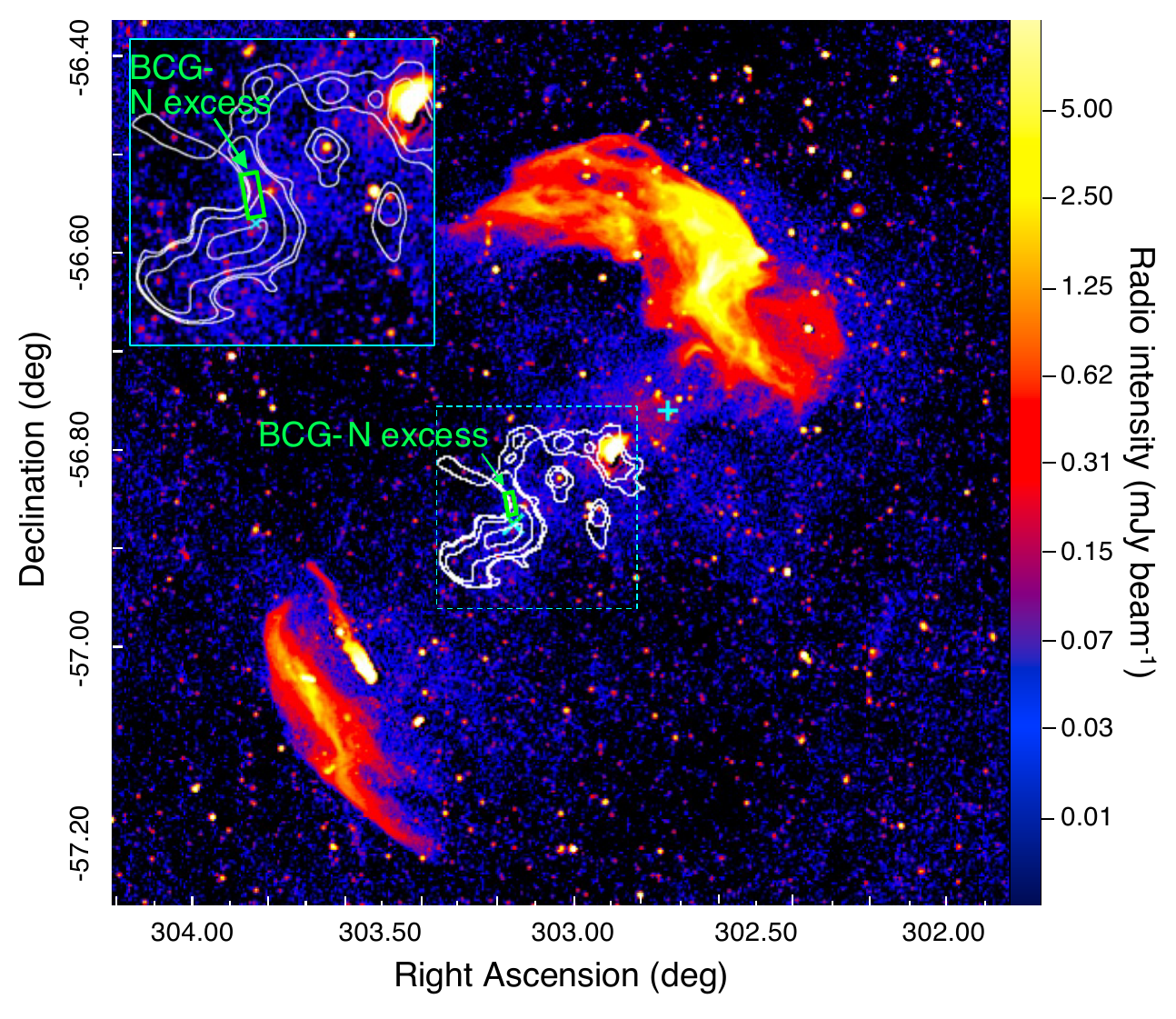}
 \end{center}
 \caption{Radio total intensity map with the ASKAP. Cyan crosses indicate the locations of the first BCG and the second BCG, individually. The white contours are the same as the black ones in Fig.~\ref{fig3.1_Abell3667_image}. The inset panel shows the enlarged image of the dotted cyan rectangle region. The cyan rectangle indicates the radio intensity excess region in the BCG-N tail.}
 \label{fig6.1_ASKAP_image}
\end{figure}

In Figure~\ref{fig6.1_ASKAP_image}, we compared the EMU ASKAP archival image of Abell~3667 at 943~MHz with our X-ray data. Interestingly, the excess of radio intensity is seen in the northern direction of the BCG. Within the $2.5\times 1.0$~arcmin$^2$ region indicated by the cyan rectangle in the inset panel of Fig.~\ref{fig6.1_ASKAP_image}, the average intensity is 76~$\mu$Jy (3.8$\sigma$ with respect to the rms noise), which is between three and ten times higher than that in the surrounding regions. Its location is consistent with that of the BCG-N tail.
Similar enhancement is also visible in the MeerKAT image at 1.2~GHz, with lower angular resolution of $35''\times 35''$\citep{2022A&A...659A.146D}.
If the offset merger scenario is correct, it is probably evidence of particle acceleration and magnetic field enhancement by turbulence generated by the backflow of the moving subcluster core into the first BCG.
A detailed multiwavelength radio study will be performed by the existing ASKAP EMU project team (Riseley et al. in preparation).

 \section{Conclusions}

In this work, we analyzed the XMM--{\it Newton} image and spectra to study the ICM distribution in Abell~3667.
In addition,
we calibrated the EPIC-PN energy scale
by following the approach proposed by \citet{2020A&A...633A..42S}
and obtained an accurate measurement of the ICM bulk-velocity distribution.

In our imaging analysis, we point out the ``RG1 vortex'' for the first time, a large clockwise vortex enhancement with a radius of about 250~kpc, connecting the first BCG and the RG1.
The RG1 vortex has a metal abundance of $\sim$0.5 solar, which is comparable to the cluster central region and about twice that in the neighboring regions.
This result suggests that the RG1 vortex is a remnant of the cluster core.
We discuss the possibility that the RG1 vortex is a Karman vortex. While feasible, detailed fluid dynamics simulations are required to decipher whether or not such a large vortex can form within the merger, which is beyond the scope of this paper.

Another potent scenario is the offset merger case. If there were even a small offset between the two cluster cores when they cross their nearest point, it is quite plausible that the mutually stripped ICM cores would have sufficient angular momentum to produce vortex-like structures.

The newly obtained results from the line-of-sight ICM velocity measurements, employing the gain recalibration method, show that region~6, which includes the BCG-N tail associated with the RG1 vortex, exhibits a redshift of 980$\pm$440~km~s$^{-1}$ compared to the optical redshift of the first BCG. 
The marginal dispersion in the line-of-sight velocity of the ICM can be explained if the off-set distance has some line-of-sight component.
At this stage, however, the observational redshift differences are not precise enough to firmly confirm this scenario. Also, better velocity measurement is needed to discuss the overall merger geometry, highlighting the necessity for {\it XRISM} mapping observations.
We also find that the ASKAP data at 943~MHz show an excess of radio intensity near the BCG-N tail, which is also evident in the MeerKAT data \citep{2022A&A...659A.146D}, indicating enhancement in either magnetic field or relativistic electron energy density in the region.

\begin{acknowledgements}
This work is supported in part by JSPS KAKENHI Grant Numbers JP15H03639, JP20H00157 (K.N.), JP22K18277, JP20K20527, JP20K14524 (Y.I.), and JP21H01135 (T.A.), and by JSPS Core-to-Core Program Grant Numbers JPJSCCA20200002 (KMI) and JPJSCCA20220002 (XRISM core-to-core). The author, Y.O., would like to take this opportunity to thank the ``Nagoya University Interdisciplinary Frontier Fellowship'' supported by Nagoya University and JST, the establishment of university fellowships towards the creation of science technology innovation, Grant Number JPMJFS2120. This work made use of data from the Galaxy Cluster Merger Catalog (\url{http://gcmc.hub.yt}).
This scientific work uses data obtained from Inyarrimanha Ilgari Bundara / the Murchison Radio-astronomy Observatory. We acknowledge the Wajarri Yamaji People as the Traditional Owners and native title holders of the Observatory site. CSIRO’s ASKAP radio telescope is part of the Australia Telescope National Facility (https://ror.org/05qajvd42). Operation of ASKAP is funded by the Australian Government with support from the National Collaborative Research Infrastructure Strategy. ASKAP uses the resources of the Pawsey Supercomputing Research Centre. Establishment of ASKAP, Inyarrimanha Ilgari Bundara, the CSIRO Murchison Radio-astronomy Observatory and the Pawsey Supercomputing Research Centre are initiatives of the Australian Government, with support from the Government of Western Australia and the Science and Industry Endowment Fund.
\end{acknowledgements}

\bibliographystyle{aa} 
\bibliography{aa} 

\begin{appendix}
\section{Detail of the measuring bulk-velocity procedure}

\label{section:SD20}
\begin{figure}[htp]
 \begin{center}
  \includegraphics[width=7cm]{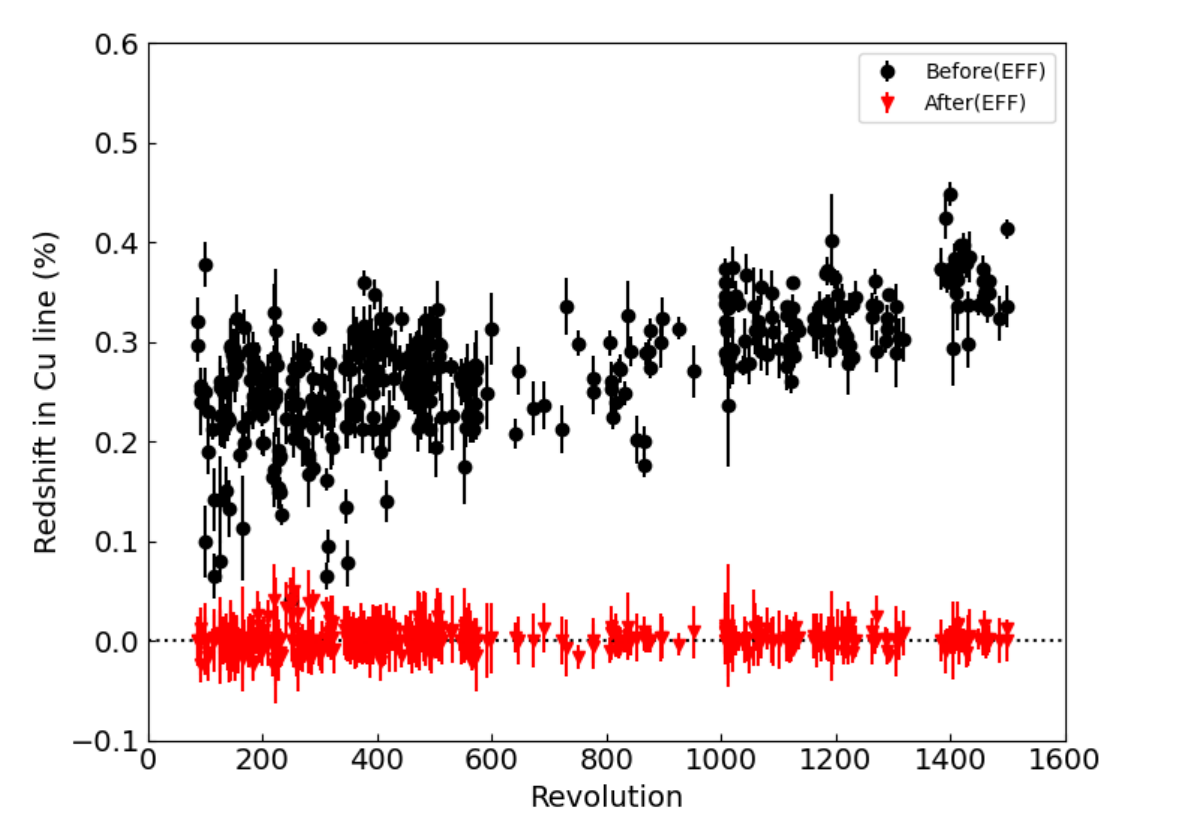}
 \end{center}
 \caption{The redshift in the Cu--K$\alpha$ line as a function of time in revolutions from 0--1500. One revolution is $\sim$2 days. The redshift of 0.1\% is about 8~eV in the Cu-K$\alpha$ line. The red markers indicate the results of the re-measurements after the first-order correction, which corrects for these redshifts by multiplying PI by 1+$z$.}
 \label{fig1_first_order_correction}
\end{figure}

\begin{figure*}[h!]
 \begin{center}
  \includegraphics[width=14cm]{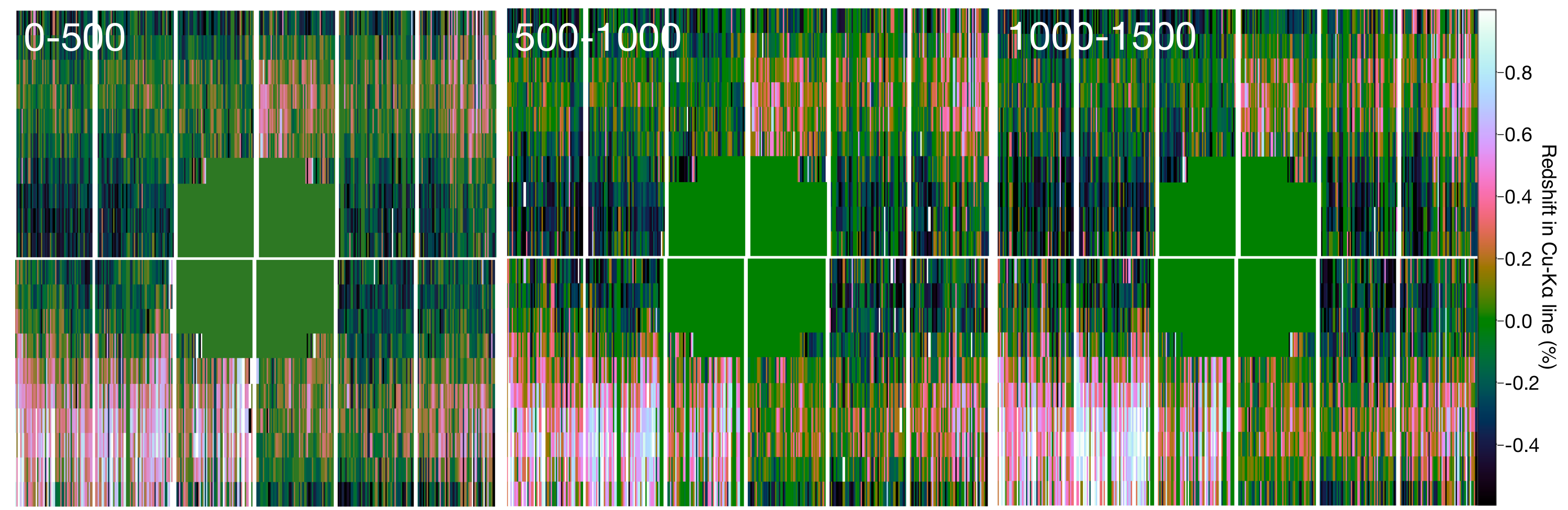}
 \end{center}
 \caption{The redshift (in percent) in the Cu--K$\alpha$ line as a function of detector position in every 500 revolutions (0-500, 500-1000,1000-1500) after the first-order correction. The spectra fitting was performed in each 1$\times$20~pixel region.}
 \label{fig2_second_order_correction}
\end{figure*}

The data with the EFF mode underwent filtering to ensure a minimum count of 10,000 between 7.0 and 9.25 keV, aiming to minimize statistical errors in individual data points.
Using this dataset, we estimated the gain re-calibration function in three steps of refinement. In the first-order correction, the redshift ($z$) of the Cu--K$\alpha$ line was measured in each observation and corrected to zero. The spectrum was extracted from a spatial region excluding the ``Cu-hole'', a polygon region at the center of the detector defined by \cite{2020A&A...633A..42S}. We fitted the spectrum in the energy range 7.0--9.25~keV with a power-law model and four Gaussian models (Ni--K$\alpha$, Cu--K$\alpha$, Zn--K$\alpha$, Cu--K$\beta$). The photon index of the power-law model was set to 0.136, an average value in \citet{2020A&A...633A..42S}. We utilized a response matrix file with an energy bin of 0.3~eV created by the {\it rmfgen} task, without utilizing an ancillary response file. The line energy positions of the Gaussian components were left unconstrained, while the sigmas were fixed at 0.0. The spectrum fitting was done by C-statistic in {\it xspec}.

The redshift of the Cu--K$\alpha$ line as a function of time in each observation is shown in Fig.~\ref{fig1_first_order_correction}. A 0.1 percent corresponds to $\sim$8~eV. We calibrated the gain shift by multiplying the PI value by 1+$z$. The results of the redshift re-measurements after first-order correction are also shown in Fig.~\ref{fig1_first_order_correction}. Note that the similar long-term CTI using the Cu--K$\alpha$ line was already updated in May 2023 (XMM-CCF-REL-389), but we have not adopted it.

In the second-order correction, the redshift of the Cu--K$\alpha$ line depending on the detector position was measured and corrected. The first-order corrected data were stacked every 500 revolutions (i.e. divided into three time-dependent groups). The spectra in the stacked data were extracted from 1~$\times$~20 pixel regions in each CCD. We fitted the spectra in 7.7--8.3~keV with a power-law and a Gaussian model (Cu--K$\alpha$). The fitting model is the same as that used in the first-order correction. The redshift maps in Cu--K$\alpha$ in each fine region are shown in Fig.~\ref{fig2_second_order_correction}.
Although the region size is slightly different from that of \cite{2020A&A...633A..42S}, our results are remarkably similar to those shown in Fig.~6 of \cite{2020A&A...633A..42S}. In the standard calibration, the Mn--K$\alpha$ and Al--K$\alpha$ emission lines in the CAL-FWC data are used to correct the gain along the readout direction at each CCD node. There is a nonuniformity in the illumination intensity of the calibration source, which is strong between CCD4 and CCD7. The intensity around the readout nodes of CCD7 and CCD9--12 is about 200 times weaker than this region, so the calibration functions at these positions have a large redshift or blue shift in Cu-K$\alpha$. We also calibrated the stacked data by multiplying the PI value by 1+$z$.

\begin{figure}[hbp]
 \begin{center}
  \includegraphics[width=7cm]{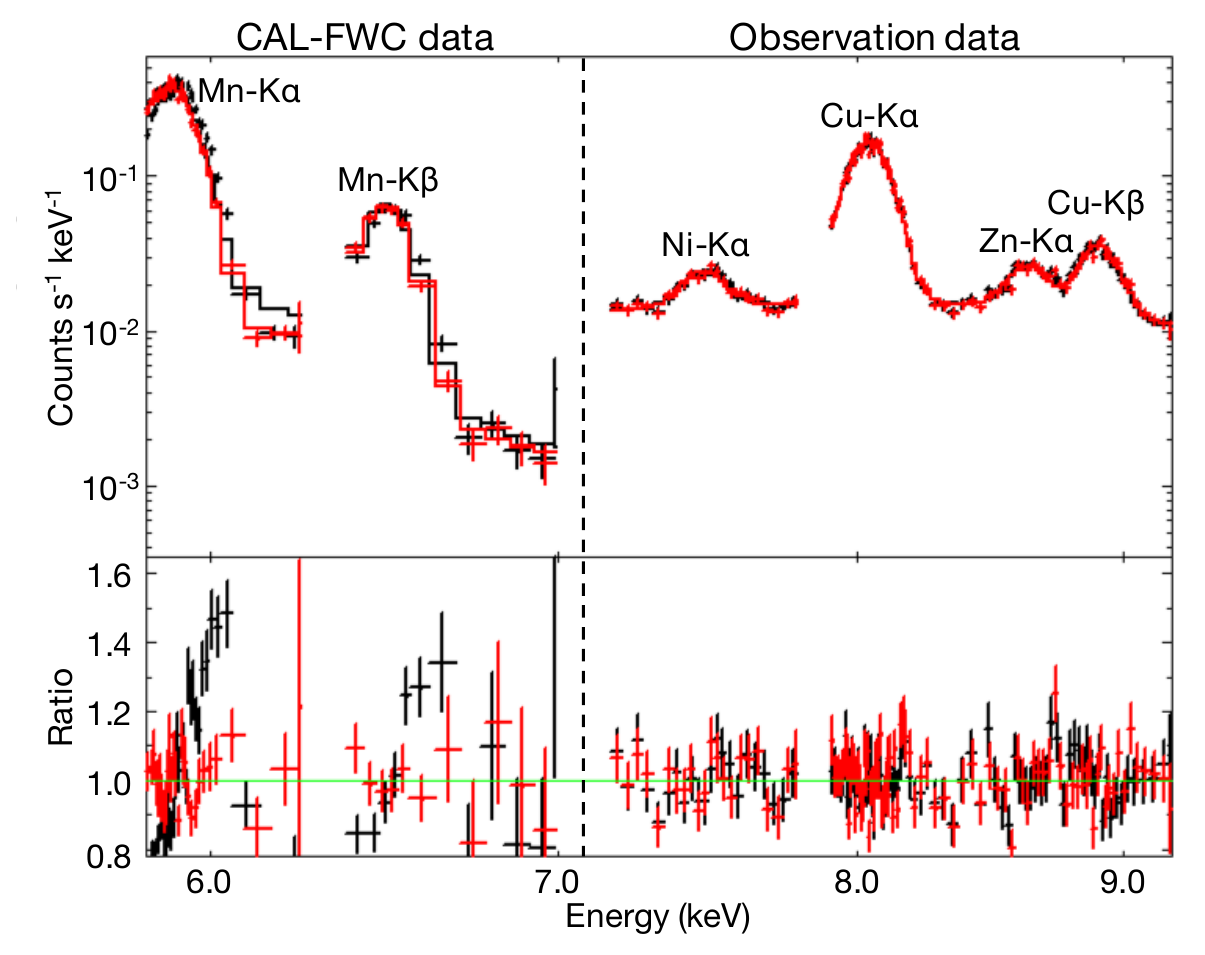}
 \end{center}
 \caption{An example of the spectrum in the CAL-FWC data and the observation data after the second-order correction. The position of the emission line in the fitting model is fixed at the theoretical value. Red points indicate spectra with the third-order correction applied, in which the energy scale deviations are corrected by a compression model based on the Cu-K$\alpha$ line.}
 \label{fig2.3_third_order_correction}
\end{figure}

In the third-order correction, the energy scale was corrected using several emission lines in the observational data and CAL-FWC data, incorporating the first-order and second-order corrections. In this turn, we defined 8 regions in each CCD, divided into four segments in the RAWX direction (0--16,17--32,33--48,49--64) and two segments in the RAWY direction (0-110,110--200). In spectral fitting, we considered 6 emission line models consisting of Mn--K$\alpha$, Mn-K$\beta$, Ni-K$\alpha$, Cu-K$\alpha$, Zn-K$\alpha$, and Cu-K$\beta$ in the 5.825--6.245, 6.375-7.000, 7.170-7.795, and 7.905--9.200 keV energy bands. As an example, Fig.~\ref{fig2.3_third_order_correction} presents a spectrum in a region of CCD3 in the 0-500 revolution. The black model shows the fitting result with the emission line position fixed at the theoretical value for comparison. Based on the Cu-K$\alpha$ emission lines, it seems that the low energy emission lines shift to the lower energy and the high energy emission lines shift to the higher energy. Therefore, we constructed a gain compression model based on the Cu-K$\alpha$ emission line as described in \cite{2020A&A...633A..42S}. For the data points of the emission line positions obtained by the fit ($E$) versus the theoretical emission line positions ($E'$), we conducted a chi-squared fit weighted by the error of each emission line, using a model of $E'=E(1+A(E-8.04))$. The resulting correction function was used to calibrate the energy scale in each region. The red line in Fig.~\ref{fig2.3_third_order_correction} shows the result of the spectrum fitting after the third-order correction.

In spectral analysis, we selected single events (PATTERN==0) for a good energy resolution and flagged data (FLAG==0) to remove events close to bad pixels. The response matrix file is with a 0.3~eV bin generated using {\it rmfgen} task. The ancillary response file for a source model is created using {\it arfgen} task with the parameter extendedsource=extend. The background components are reproduced by a power-law model with a photon index of 0.136 and four Gaussian models as well as the first-order correction. We adopt the ``total fitting technique'' described in \citet{2020A&A...633A..42S}.  Specifically, the spectra were added together using {\it mathpha} and the response files were added together with a weighting factor of the numbers of counts in the 4--9.25~keV energy band using {\it addrmf} and {\it addarf}. Velocity measurements in the Coma cluster, adapting the estimated energy scale recalibration function, are in good agreement with the result of \citet{2020A&A...633A..42S}. The details are described in Appendix~\ref{appendix:Coma}.

\section{Cross-check using the Coma cluster}
\label{appendix:Coma}

\begin{figure}[htp]
 \begin{center}
  \includegraphics[width=7cm]{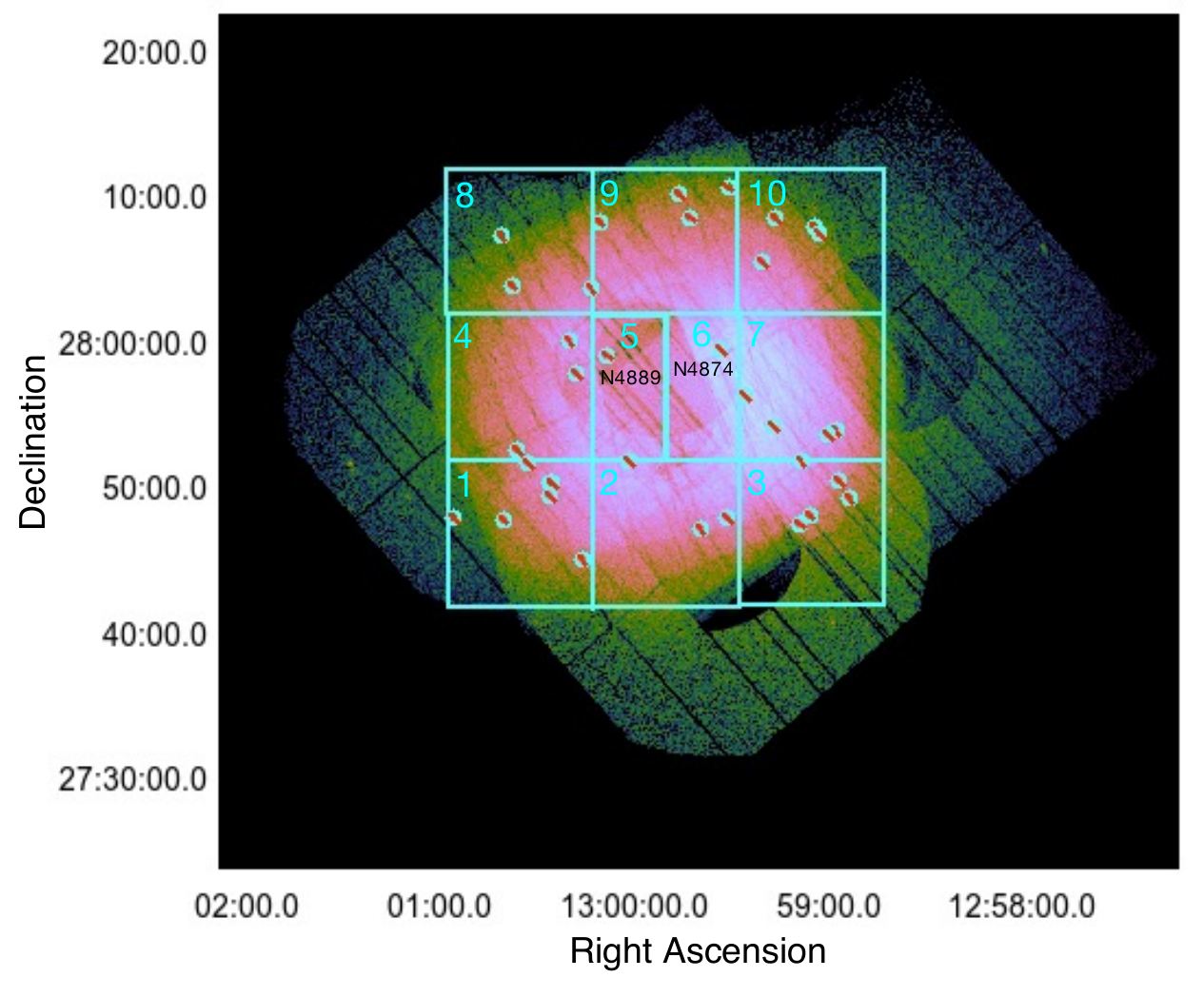}
 \end{center}
 \caption{The mosaic count image without the Cu hole in the Coma cluster. The energy range is 0.5-10.0~keV. The rectangles in cyan show the examined regions, which were divided according to \citet{2020A&A...633A..42S}. The circles indicate the point sources excluded in our analysis.}
 \label{fig2.4_coma_mosaic_image}
\end{figure}

For comparison with the results of \citet{2020A&A...633A..42S}, we have measured the bulk-velocity structure in the Coma cluster using our calibration function. The data used are listed in table~\ref{table1_Coma_obsid}. We used the data taken in the EFF mode. Therefore the observation 0153750101 was excluded.  As mentioned in the data reduction, we extracted the event files in each observation using the {\it epchain} task in SAS and filtered out the soft proton effect. The events outside the Cu-hole are re-calibrated by taking 3~steps. In the first-order correction, we used the exposure time weighted redshift of the period around 100 revolutions based on the observational data. A mosaic image in the Coma cluster was shown in Fig.~\ref{fig2.4_coma_mosaic_image}. The hydrogen column density was fixed at $8.54\times10^{19}~\rm{cm^{-2}}$, determined from the HI map.

\begin{figure*}[htpb]
 \begin{center}
  \includegraphics[width=12cm]{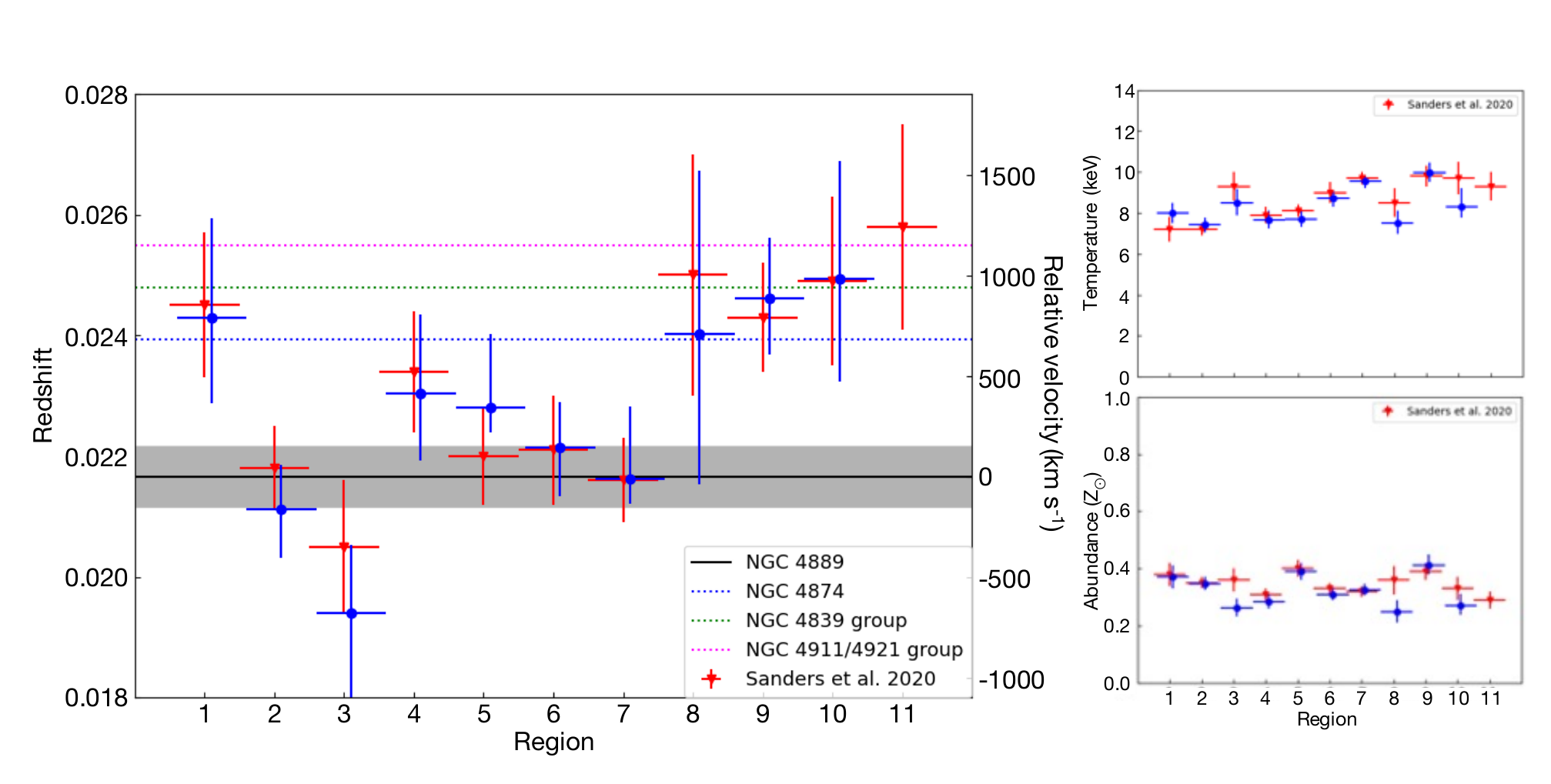}
 \end{center}
 \caption{The result of redshift (left), temperature (lower-right), and abundance (upper-right) in the Coma cluster, comparing the result of \citet{2020A&A...633A..42S}. The right axis in the redshift distribution shows relative velocity based on the redshift of NGC~4889 (z=0.0217).}
 \label{fig2.2_coma_result}
\end{figure*}

A mosaic count image without the Cu hole in the Coma cluster is shown in Fig.~\ref{fig2.4_coma_mosaic_image}. As with the \citet{2020A&A...633A..42S} results, we divided the Coma cluster into 10 regions as shown in Fig.~\ref{fig2.4_coma_mosaic_image}. In the spectra analysis we excluded the point sources, which are shown as circles with a radius of 30~arcsec. Figure~\ref{fig2.2_coma_result} shows the bulk-velocity distribution using our calibration function, overlapping the results of \citet{2020A&A...633A..42S}. Our results reproduce their results with good reproducibility.

\begin{table}
\label{table1_Coma_obsid}
\caption{The observation list in the Coma cluster.}
\centering
\begin{tabular}{c|c|c|c}
    \hline \hline
    obsid & Duration & Revolution & mode \\
    \hline \hline
    0124710501 & 30~ks & 86 & EFF \\
    0124710601 & 32~ks & 93 & EFF \\
    0124710901 & 31~ks & 93 & EFF \\
    0124711401 & 35~ks & 86 & EFF \\
    0124712001 & 23~ks & 184 & EFF \\
    0300530101 & 26~ks & 1012 & EFF \\
    0300530201 & 28~ks & 1011 & EFF \\
    0300530301 & 31~ks & 1008 & EFF \\
    0300530401 & 28~ks & 1007 & EFF \\
    0300530601 & 26~ks & 1006 & EFF \\
    0300530701 & 26~ks & 1006 & EFF \\
    \hline
\end{tabular}
\end{table}

\section{Detail of the offset merger simulation}
\label{Detail of the offset merger simulation}

\begin{figure*}[htpb]
 \begin{center}
  \includegraphics[width=13cm]{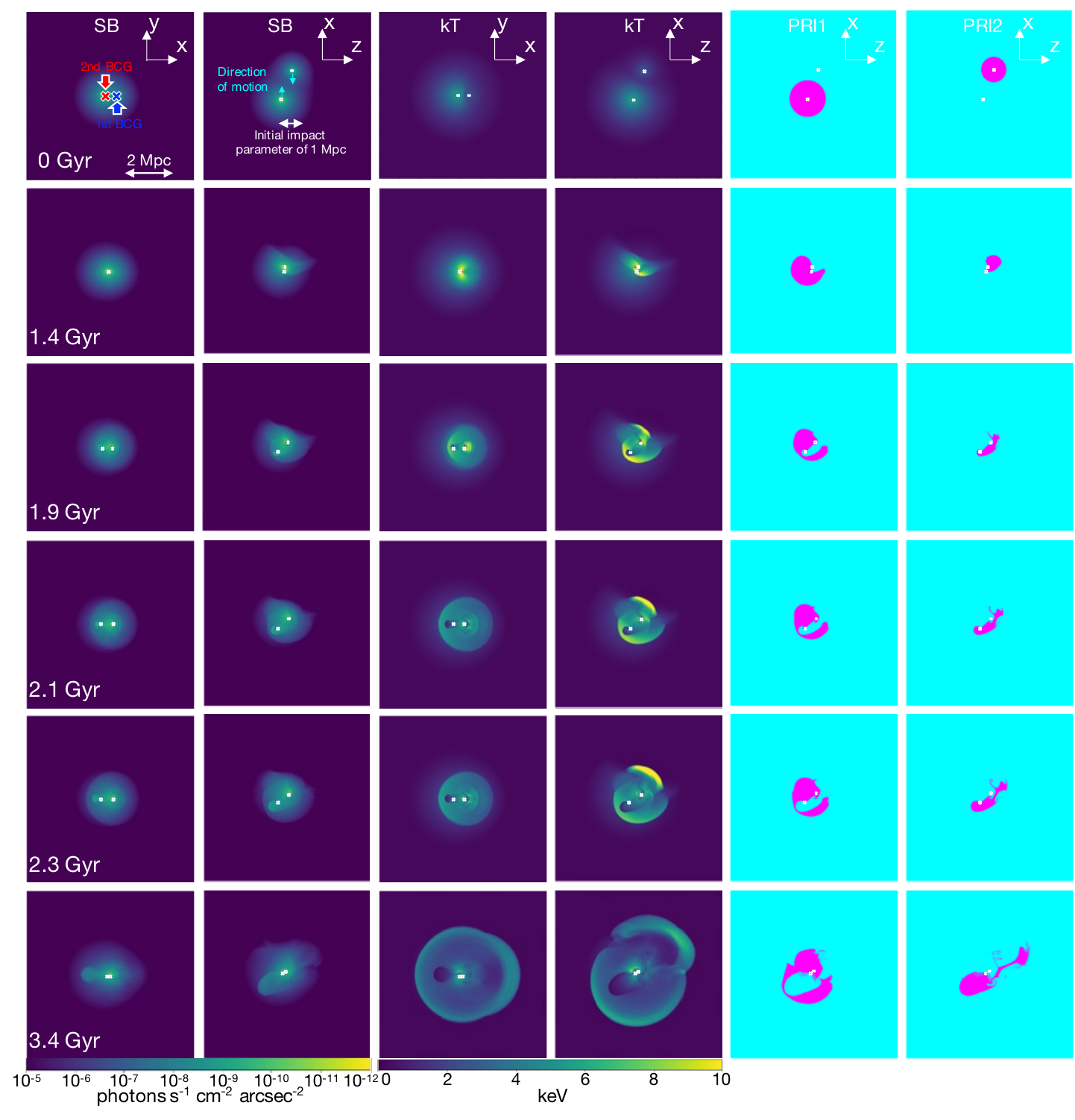}
 \end{center}
 \caption{The simulation results of the merging cluster with an initial parameter of mass ratio of 3:1 in each epoch (0~Gyr, 1.4~Gyr, 1.9~Gyr, 2.3~Gyr, and 3.4~Gyr) provided by the Galaxy Cluster Merger Catalog \citep{2016arXiv160904121Z,2018ApJS..234....4Z}. The initial impact parameter is set to 1000~kpc in the z-axis direction, and the two clusters approach each other by moving along the x-axis. From the left panel to the right panel: X-ray surface brightness observed from the +z-axis direction, the +y-axis direction, the projected temperature observed from the +z-axis direction, the +y-axis direction, and the projected regions of the ICM of the main cluster (PRI1) and subcluster (PRI2) from the +y-axis direction, respectively. The position of the BCG in each cluster is marked with blue and red crosses.}
 \label{fig3.3_zuhone_simulation}
\end{figure*}

Figure~\ref{fig3.3_zuhone_simulation} shows the result of a 3:1 mass cluster merger simulation with an initial impact parameter of 1000~kpc provided by the Galaxy Cluster Merger Catalog \citep{2016arXiv160904121Z,2018ApJS..234....4Z}, as noted in Sec.~\ref{Comparison with simulation}.

\end{appendix}

\end{document}